\documentclass[11pt]{article}
\usepackage{a4,exscale,epsf,graphics}
\def\Re{{\rm Re}}
\def\Im{{\rm Im}}
\begin{document}
\def\IncludeEpsImg#1#2#3#4{\renewcommand{\epsfsize}[2]{#3##1}{\epsfbox{#4}}}
\def\fsz{\footnotesize}
\def\Pade{Pad\'e}
\begin{center}
\textbf{\Large Fast-Convergent Resummation Algorithm and
 Critical Exponents of $\phi^4$-Theory in Three
  Dimensions}
\vskip 0.4in
 \textbf{Florian Jasch} and \textbf{Hagen Kleinert}
\footnote{Emails: jasch@physik.fu-berlin.de~kleinert@physik.fu-berlin.de ~~~~~~~~~~~~~~~~~
~~~~ \hfill\linebreak URL:
http://www.physik.fu-berlin.de/\~{}kleinert~, Phone/Fax:
 0049 30 8383034.}
\vskip0.5cm
~Institut f\"ur Theoretische Physik \\
Freie Universit\"at Berlin, Arnimallee 14, 1000 Berlin 33, Germany
\end{center}
\begin{abstract}
We develop an efficient algorithm for
evaluating divergent
perturbation expansions of field theories
in the bare coupling constant  $g_B$
for which we possess a finite number $L$ of expansion coefficients
plus two more informations:
The  knowledge of the
large-order behavior proportional to $(- \alpha )^kk!k^\beta g_B^k$,
with a known growth parameter $ \alpha $,
and the knowledge of the
approach to scaling
 being
of the type $c+c'/g_B^ \omega $, with constants $c,c'$ and a
critical exponent of
approach $ \omega $.
The latter  information
leads to an increase
in the speed of convergence
and a high accuracy of the results.
The algorithm is  applied
to the six- and seven-loop
expansions for the critical exponents of O($N$)-symmetric
$\phi^4$-theories, and the result for the critical exponent
$ \alpha $ is compared with the recent satellite
experiment.
\end{abstract}
\baselineskip 0.75cm
\def\betaj{  \beta _0 }
\def\deltaj{ \delta }
\def\sigmaOj{\alpha}
\def\sigmaj{\sigma}
\def\bOj{\beta}
\def\beq{\begin{equation}}
\def\eeq{\end{equation}}
\def\ins#1{}
\def\ep{\varepsilon}
\def\lfrac#1#2{{#1/#2}}
\def\sbf#1{\mbox{{\scriptsize$\bf{#1}$}}}

\section{Introduction}
The field-theoretic approach to critical phenomena provides us with power
series expansions
for the
critical exponents of a wide variety of universality classes.
For $\phi^4$-theories with O($N$) symmetry
in three dimensions,
these expansions have been calculated
numerically as power series in the renormalized coupling constant
up to seven loops for the critical exponents $ \nu $ and $ \eta $
 \cite{nickel} and up to six loops for
the exponent $ \omega $ governing the approach to scaling.
In $4-\epsilon$ dimensions, exact $\epsilon$-expansions are available
up to five loops for all critical exponents
with  O($N$) symmetry \cite{KNS}, cubic symmetry, and mixtures of these
\cite{KSF}.
When inserted into the renormalization group equations,
these expansions are supposed to determine
the critical  exponents via their values
at an infrared-stable fixed point $g=g^*$.
The latter step is nontrivial since
the expansions are divergent and require resummation,
for which sophisticated methods have been developed,
summarized and applied most recently in  \cite{GZ}.
The resummation methods use the information from the known large-order
behavior
 $(-\alpha)^kk!k^\beta g_B^k$
of the
expansions and analytic mapping techniques to obtain quite accurate results.

A completely different resummation procedure
was developed recently
on the basis of  variational perturbation theory \cite{PI}
 to the expansions in powers of the {\em bare coupling constant\/},
which goes to infinity at the critical point.
The resulting {\em strong-coupling theory\/} \cite{scth}
was successfully applied in three \cite{PI3,seven} and
$4-\epsilon$ dimensions \cite{PIep},
and rendered for the first time an interpolation between expansions of $4-\epsilon$
and $2+\epsilon$-dimensional theories.
This method converges as fast as the previous ones,
even though it does not take into
account the information on the large-order behavior of the expansions.
Instead, it uses the fact that the power series for the critical exponents
approach their constant critical
value in the form $c+c'/g_B^ \omega $,
where $c,c'$ are constants, and
$ \omega $ is the
critical exponent of the approach to scaling.
The results showed that the latter  information
is just as efficient in increasing the speed of convergence
as the information on the large-order behavior.

We may therefore expect that a resummation method
which incorporates both  informations
should lead to results with an even higher accuracy,
and
it is the purpose of this paper to present such a method in the form
of a simple algorithm.

\section{The Problem}
The development
of our resummation algorithm
is based on an improvement
of the problem formulated
in \cite{scth,z1112}
and solved via variational perturbation theory \cite{PI}.
Mathematically, the problem we
want to solve is the following:
Let
\begin{equation}
  f_L(g_B){=}\sum_{k=0}^L f_k g_B^k
\label{asym}
\end{equation}
be
the first $L$ terms
of a divergent
asymptotic expansion
\begin{equation}
  f(g_B){=}\sum_{k=0}^\infty f_k g_B^k
\label{asymall}
\end{equation}
of a function $f(g_B)$,
 which possesses a
strong-coupling expansion of the type
\begin{equation}
\label{stark}
f(g_B)=g_B^{s}\sum_{k=0}^\infty b_k g_B^{-k\omega},
\end{equation}
which is assumed to have some finite convergence radius $|g_B|\ge g_B^{\rm conv}$.
Suppose that the function is
analytic
in the complex $g_B$-plane
with a cut along the negative real axis, with a
discontinuity  known from
instanton calculations \cite{parisi2,ch17}
to have near the tip of the cut the generic form
\begin{equation}
{\rm disc}f(-g_B)\equiv 2\pi i\, \gamma \,( \alpha|g_B|)^{-\beta-1}e^{-1/\alpha|g_B|}.
\label{@}\end{equation}
Via a dispersion relation,
\begin{equation}
f(g_B)=\frac{1}{2\pi i}\int _{0}^{\infty}dg'_B\frac{{\rm disc}f(-g_B)}{g_B'+g_B},
\label{eq:imp}\end{equation}
or a sufficiently subtracted version of it,
this discontinuity
corresponds to the
large-order behavior
of the expansion coefficients $f_k$
\begin{equation}
  \label{eq:wachs}
  f_k\stackrel{k \to \infty}{=}\gamma k! (-\alpha)^k k^{\beta}
  [1+\mathcal{O}(1/k)].
\end{equation}
The constant $\alpha$ is given by the inverse action of the radial symmetric
solution
to the classical field equations. The parameter $\beta$
counts the number of zero modes
in the fluctuation determinant around this solution. The absolute
normalization $\gamma$ of the large-order behavior requires the calculation of
the fluctuation determinant \cite{parisi2}.

As far as the leading
 strong-coupling coefficient $b_0$ is concerned,
this problem has been attacked before by
Parisi \cite{parisi} using
a resummation method based on
Borel transformations in combination with analytic mapping techniques.
However, when applied to the asymptotic
expansions of the ground state energy of
the anharmonic oscillator, his  method
converged very slowly,  too slow to
lead to reliable critical
exponents, where only five to seven expansion coefficients $f_k$ are known.
The reason is that in Parisi's approach, the
corrections to the leading power behavior are
failing to match the true fractional powers of the
strong-coupling expansion (\ref{stark}).

This deficiency was cured by the strong-coupling
theory of one of the authors (HK) in Ref.~\cite{scth},
and the subsequent application to critical exponents
in Refs.~\cite{PI3,seven,PIep}, which
showed a surprisingly rapid convergence.
However, that theory did not take advantage
of the knowledge of the large-order behavior (\ref{eq:wachs}),
which can lead to
an increase in the speed of convergence and thus
of the accuracy of theoretical values for the critical exponents.
This will be done in the present improved resummation method.

\section{Borel Methods}

Basis for this method
is the development of a
 more general Borel-like transformation
which will automatically guarantee
the form of the strong-coupling expansion (\ref{stark})
for each
approximant $f_L(g_B)$.
Let us first
recall
briefly the
important properties of the ordinary Borel transformations:
It is a function $B(t)$ associated
with $f(g_B)$ which is defined
by the
Taylor series
\begin{equation}
  \label{eq:boreltrans}
  B(t)=
 \sum_{k=0}^\infty B_k t^k\equiv
 \sum_{k=0}^\infty \frac{f_k}{k!} t^k.
\end{equation}
By dividing the expansion coefficients $f_k$
by $k!$, the factorial growth of
$f_k$ is reduced to a power growth, thus giving $B(t)$
a finite convergence radius.

An alternative definition of the Borel transform is
given by the
contour integral
\begin{equation}
  \label{eq:intdarstborel}
  B(t)\equiv \frac{1}{2\pi i}\oint_C\frac{dz}{z}e^z \; f(t/z),
\end{equation}
where the contour $C$ encloses anticlockwise the negative real axis.
Indeed, inserting here (\ref{asym}) and performing the
integral we obtain (\ref{eq:boreltrans}).

If $f(g_B)$ is an analytic function in the sector
\begin{equation}
  \label{eq:sektor}
  S_{\pi/2+\delta}^R\equiv \{g_B\;|\;|g_B|<R,|\arg(g_B)|<\pi/2+\delta \}
\end{equation}
of a circle, and satisfies the so-called strong asymptotic condition
\begin{equation}
  \label{starkasym}
|f(g_B)-\sum_{k=0}^Lf_kg_B^k|<A g_B^{L+1}\alpha^{L+1}(L+1)!~,\hspace{1cm}
{\rm with}~~\alpha,A >0,
\end{equation}
then $B(t)$ is analytic in $S_{\delta}^{\infty}$,
with a finite
 radius of convergence $t<1/\alpha$.
 The original function
$f(g_B)$ can be recovered from $B(t)$ by the inverse Borel transformation
\begin{equation}
  \label{eq:umkehrtrans}
  f(g_B)=\int_0^{\infty}e^{-t}B(tg_B)dt.
\end{equation}
Obviously, the inverse transformation can only be performed
if $B(t)$
is known on the entire
 positive real $t$-axis.
The
 Taylor series (\ref{eq:boreltrans}) for $B(t)$,
however, converges only
inside the circle of radius $1/ \alpha$. Before we can do the integral in
(\ref{eq:umkehrtrans}), we must therefore
 perform a suitable analytic
 continuation of (\ref{eq:boreltrans}) \cite{z13}.
This can be done by reexpanding  $B(t)$ in powers of the
function $\kappa(t)$
defined implicitly by
\begin{equation}
  \label{eq:konfab}
  t=\frac{1}{\sigma}\frac{\kappa(t)}{[1-\kappa(t)]^p}.
\end{equation}
This function maps the interval $[0,\infty]$
of the $t$-axis to the interval $[0,1]$ of the $\kappa$-plane. By a
proper choice of $\sigma$ it is possible to make the  unit circle
free  of singularities.
Then we may
use the reexpansion $B(t)$ in powers of $\kappa(t)$
 truncated after $ \kappa ^L$,
\begin{equation}
B_L(t)\equiv \sum _{k=0}^L v_k  \kappa^k (t),
\label{@reexpansi}
\end{equation}
as an approximation to $B_L(t)$ on the entire positive real $t$ axis.
Inserting this into the
inverse transformation formula (\ref{eq:umkehrtrans}), we obtain
an approximation $f^a_L(g_B)$  for $f(g_B)$, which has the same first $L$
expansion
coefficients as $f_L(g_B)$
and, in addition, the correct large-order behavior (\ref{eq:wachs}).

How can we incorporate the strong-coupling expansion (\ref{stark})
of $f(g_B)$ into the approximation $f_L^a(g_B)$?
In the Borel transform $B(t)$, the
 strong-coupling expansion (\ref{stark})
amounts to a large-$t$ expansion
\begin{equation}
  \label{eq:starkborel}
  B(t) = t^s\sum_{k=0}^\infty\frac{\sin\pi(k\omega-s)}{\pi}
  \Gamma(k\omega-s)b_kt^{-k\omega}  .
\end{equation}
This follows directly
by inserting (\ref{stark}) into
(\ref{eq:intdarstborel})
and integrating each term.
Here and in the sequel,
$C$ denotes a path of integration which encloses anticlockwise the
negative real axis in the complex plane.

If the series (\ref{stark}) has a finite radius
of convergence,
the large-$t$ expansion of $B(t)$
is a divergent
asymptotic one, because  of the factor $ \Gamma (k \omega -s)$ in
the $k$-th expansion
coefficient.

It should be stressed, that the relation between the coefficients of the
strong-coupling expansion (\ref{stark}) and the coefficients of expansion
(\ref{eq:starkborel}) is not generally invertible, because of the factor $\sin\pi(k
\omega-s)$ which causes the coefficients of negative integer powers of
$t$ to vanish.

Note that, in general,
an expansion
in the Borel-plane with a power sequence in $t$ as in
(\ref{eq:starkborel})
is not sufficient to
ensure an expansion in the same powers in the $g_B$-plane
as in (\ref{stark}), because
of the appearance of extra integer powers in $g_B$.
This is illustrated by the simple function $B(t)=(1+t)^s$,
which possesses a strong-coupling expansion in the powers $t^{s-k}$.
If $s$ is non-integer the expansion of the corresponding function $f(g_B)$
reads
\begin{equation}
f(g_B)=\int_0^\infty dt\;
e^{-t}(1+g_Bt)^s =e^{1/g_B} \Gamma(s+1)g_B^s+e^{1/g_B}\sum_{k=0}^\infty
\frac{(-1)^k}{(k+s+1)k!}g_B^{-k-1},
\end{equation}
and expanding the exponential we see that the sum
contains
integer powers which are not contained in the
strong-coupling expansion of $B(t)$.

It is advantageous to perform a further analytic
continuation of the reexpansion (\ref{@reexpansi})
which enforces automatically the
leading power
behavior
$t^s$ of $B(t)$.
For this we
change
(\ref{@reexpansi})
to
\begin{equation}
B_L(t)\equiv
 [1-\kappa(t)]^{-ps}\sum_{k=0}^{L}h_k~\kappa^k(t).
\label{@inderti}\end{equation}
The coefficients
$h_k$ are determined by using (\ref{eq:konfab}) to expand
$ \kappa (t)$ in powers of $t$, inserting this into
(\ref{@inderti}), reexpanding in powers of $t$,
and comparing the final coefficients with those in (\ref{eq:boreltrans}).
When the approximation (\ref{@inderti}) is inserted
into (\ref{eq:umkehrtrans}), we obtain $f^a_L(g_B)$ with the
correct leading power behavior $g_B^s$ for large $g_B$.

Unfortunately, the simple prefactor
does not
produce the correct subleading powers $(g_B)^{s-k\omega}$
of the strong-coupling expansion (\ref{stark}),
and we have not been able to find another
 simple analytic continuation
of $B(t)$ which would achieve this.

\section{Hyper-Borel Transformation}
A solution of this problem is, however, possible
with the help of a
generalization of the Borel-Leroy transformation to
what we shall call a {\em hyper-Borel transformation\/}
\cite{tr}
\begin{equation}
\label{modbor}
\tilde{B}(y)
=\sum_{k=0}^\infty
\tilde B_ky^k,
\end{equation}
with  coefficients
\begin{equation}
\label{modbor}
\tilde B_k\equiv
\omega\frac{\Gamma\left(k(1/\omega-1)
+\betaj \right)}{\Gamma\left(k/\omega-s/\omega\right)\Gamma(\betaj )}f_k.
\end{equation}
\subsection{General Properties}
The inverse transformation
is given by the double integral
\begin{equation}
  \label{umkehrtrans}
f(g_B)=\frac{\Gamma(\betaj )}{2\pi
  i}\oint_Cdte^tt^{-\betaj }\int_0^\infty
\frac{dy}{y}\left[\frac{g_B}{yt^{(1-\omega)/\omega}}\right]^s\exp\left[\frac{yt^{(1-\omega)/\omega}}{g_B}\right]^\omega \tilde{B}(y),
\end{equation}
as can
easily shown with the help of the
integral
 representation
of the inverse Gamma function
\begin{equation}
  \label{eq:gammadarst}
  \frac{1}{\Gamma(z)}=\frac{1}{2\pi i}\int_Cdte^tt^{-z}.
\end{equation}
The transformation
possesses a free parameter $ \betaj  $ which will be used to optimize the
approximation $f_L(g_B)$ at each order $L$.
The  power $\omega$ of the strong-coupling expansion is assumed to
lie in the interval $0<\omega<1$, as it does in the upcoming physical applications.

The  hyper-Borel transformation
has the desired property of
allowing for a resummation of $f_L(g_B)$
with the full sequence of
powers of $g_B$ in the strong-coupling expansion (\ref{stark}).
To show this we first observe
that as in the ordinary Borel transform
(\ref{eq:boreltrans}),
the large-argument behavior of the Gamma function known from Stirling's
formula
\begin{equation}
\label{gammawachstum}
\Gamma(pk+q) \stackrel{k \to \infty}{=}
\sqrt{2\pi}^{1-p}p^{q-1/2} k^{-1/2+q-p/2}p^{pk}(k!)^p
[1+\mathcal{O}(1/k)],
\end{equation}
removes the factorial growth (\ref{eq:wachs}) from the
expansion coefficients $f_k$, and  leads
to  a simple power behavior of the coefficients $\tilde B_k$:
\begin{equation}
  \tilde B_k \stackrel{k \to \infty}{=} \mathrm{const.} \times
  \left[\sigmaOj\omega(1-\omega)^{1/\omega-1}\right]^k
  k^{\bOj+\betaj  +1/2+s/\omega}[1+\mathcal{O}(1/k)].
\label{@Thus o}\end{equation}
Thus our transform $\tilde{B}(y)$ shares with the
 ordinary
Borel transform $B(t)$ the property
of being  analytic at the origin.
Its radius of convergence is
determined by
the singularity on the
negative real axis at
\begin{equation}
\label{posbor}
y_s = -\frac{1}\sigma \equiv -\frac{1}{\sigmaOj}
\frac{1}{\omega(1-\omega)^{1/\omega-1}}.
\end{equation}

\subsection{Resummation Procedure}
A resummation procedure can now be set up on the basis of
the transform
$\tilde{B}(y)$ as before.
The inverse transformation (\ref{umkehrtrans}) contains an integral
over the entire positive axis, requiring
again an analytic continuation
of the Taylor expansion of $\tilde{B}(y)$
beyond the convergence radius.

The reason for introducing the
 transform $\tilde B(y)$ was to allow us to
reproduce the complete power sequence in the
strong-coupling expansion (\ref{stark}), with a leading power $g_B^s$ and
a subleading sequence of powers
$g_B^{s-k \omega }$, $k=1,2,3,\dots~$.
This is achieved by removing a factor $e^{-\rho\sigma y}$ with $\rho,\sigma
>0$ from the truncated
series  (\ref{modbor}) of our transform $\tilde{B}(y)$.
Furthermore by removing a second simple prefactor of the form
$(1+\sigma y)^{-\delta}$ we weaken the leading singularity in the
 hyper-Borel complex $y$-plane, which determines the large order behavior (\ref{eq:wachs}).
The remaining series has still a finite radius of convergence.
To achieve convergence on the entire positive $y$ axis
for which we must do the integral
(\ref{umkehrtrans}),
we reexpand the remaining series of $y$ in powers
of $ \kappa (y)$ which is related to $y$ by an equation like
Eq.~(\ref{eq:konfab}). For simplicity we choose
the parameter $p=1$, i.e.
\begin{equation}
  \label{eq:konfaby}
  y=\frac{1}{\sigma}\frac{\kappa(y)}{1-\kappa(y)},
\end{equation}
which maps
a shifted right half of the complex $y$-plane with
$\Re[y] \ge -1/2\sigma$ onto the unit circle
in the complex  $ \kappa $-plane.
Thus we reexpand
$\tilde{B}(y)$ in the following way:
\begin{equation}
  \label{eq:analytfort}
  \tilde{B}(y)\equiv \sum_{k=0}^{\infty}\tilde B_ky^k
= e^{-\rho\sigma y}[1+\sigma y]^{-\delta}\sum_{k=0}^{\infty}h_k\, \kappa ^k(y)
= e^{-\rho\sigma y}\sum_{k=0}^{\infty}h_k\frac{(\sigma
  y)^k}{(1+\sigma y)^{k+ \delta }}.
\end{equation}
The inverse  hyper-Borel transform
of $\tilde B(y)$
is now found by forming the integrals
of the expansion functions in (\ref{eq:analytfort})
\begin{equation}
\label{intdarst}
  I_n(g_B)
= \frac{\Gamma(\betaj )}{2\pi i}
             \oint_Cdte^tt^{-\betaj }\int_0^\infty\frac{dy}{y}
             \left[\frac{g_B}{yt^{1/\omega-1}}\right]^s
             \exp\left[-\frac{yt^{1/\omega-1}}{g_B}\right]^\omega
 e^{-\rho\sigma y} \frac{(\sigma y)^n}{(1+\sigma y)^{n+\delta}},
\end{equation}
so that the approximants $f^a_L(g_B)$
may be written as
\begin{equation}
  f^a_L(g_B)=\sum_{n=0}^L h_n I_n(g_B).
\label{complete lis}
\end{equation}

The same functions $ I_n(g_B)$
may be used  as basis functions for a wide variety of
divergent truncated perturbation expansions
$f_L(g_B)$.
The complete list of parameters on which they depend
reads as follows
\begin{equation}
I_n(g_B) = I_n(g_B,\omega,s,\rho,\sigmaj,\deltaj,\betaj )=
I_n(\sigmaj g_B,\omega,s,\rho,1,\deltaj,\betaj ),
\end{equation}
but in the following we shall mostly use the shorter notation
$I_n(g_B)$.
The integral representation of  $I_n(g_B)$ breaks
down at $s=n$, requiring an analytical continuation.
For the  upcoming applications
in the
large-$g_B$ regime it will
be sufficient to perform this continuation only in the convergent
strong-coupling expansion of $I_n(g_B)$.
This is obtained by
performing a Taylor series expansion of the
exponential function in
(\ref{intdarst}), which is an expansion in powers of $1/g_B^{\omega}$. After integrating
over $t$ and $y$ using (\ref{eq:gammadarst}), we obtain
an expansion
\begin{equation}
\label{instark}
  I_n(g_B) = g_B^s\sum_{k=0}^\infty b_k^{(n)}g_B^{-k\omega },
\end{equation}
which has indeed the same power sequence as
the  strong-coupling expansion (\ref{stark}) of the function $f(g_B)$
to be resummed.

The expansion coefficients are
\begin{equation}
\label{koeffstark}
  b_k^{(n)}= \frac{(-1)^k}{k!}
               \frac{\sigmaj^{s-k\omega}\Gamma(\betaj )}{\Gamma[(\omega-1)
               k+\betaj +(1/\omega-1)s]}i_k^{(n)},
\end{equation}
where $i_k^{(n)}$ denotes the integral
\begin{equation}
\label{koeffstarkin}
  i_k^{(n)}=\int_0^\infty dye^{-\rho
  y}(1+y)^{-\deltaj-n}y^{k\omega +n-s-1}.
\end{equation}
This integral is seen to coincide with the Kummer function
\begin{equation}
  U(\alpha,\gamma,z) \equiv\frac{1}{\Gamma(\alpha)}\int_0^\infty
  dye^{-zy}y^{\alpha-1}(1+y)^{\gamma-\alpha-1},
\end{equation}
in terms of which we can write
\begin{equation}
i_k^{(n)}=\Gamma(k\omega +n-s)\,U(k\omega +n-s,k\omega -s-\deltaj+1,\rho).
\label{@formulaU}
\end{equation}
The latter expression is useful since in some applications
the integral (\ref{koeffstarkin})
may diverge, and  requires an  analytic continuation by deforming the
contour of integration.
Such deformations are automatically supplied by choosing
other representations
for the Kummer function, for instance
\begin{equation}
  U(\alpha,\gamma,z)=\frac{\pi}{\sin \pi\gamma}\left[\frac{M(\alpha,\gamma,z)}{\Gamma(1+\alpha-\gamma)\Gamma(\gamma)}-z^{1-\gamma}\frac{M(1+\alpha-\gamma,2-\gamma,z)}{\Gamma(\alpha)\Gamma(2-\gamma)}\right],
\label{@formulaUM}\end{equation}
where $M(\alpha,\gamma,z)$ is the confluent hypergeometric function
with a Taylor expansion
\begin{equation}
  M(\alpha,\gamma,z)=1+\frac{\alpha}{\gamma}\frac{z}{1!}+\frac{\alpha(\alpha-1)}{\gamma(\gamma-1)}\frac{z^2}{2!}+\ldots.
\label{@formulaM}\end{equation}
The alternative expression  (\ref{@formulaU}) for $i_k^{(n)}$,
with (\ref{@formulaUM}) and (\ref{@formulaM}),
  is useful for resumming
various asymptotic expansions, for example that
of the ground state energy of the
anharmonic oscillator, in which case the leading strong-coupling power $s$
has the value
 $1/3$.
There the integral representation (\ref{koeffstarkin})
would have  to be evaluated for values
$n=0,~k=0$, where the integral does not exist,
whereas formula (\ref{@formulaU})
with (\ref{@formulaUM}) and (\ref{@formulaM})
is well-defined.

For large $k$, the integral on the right-hand side of (\ref{koeffstarkin}) can
be
estimated with the help of the saddle point approximation. The saddle point
lies at
\begin{equation}
  \label{eq:saddle}
  y_s\approx\frac{k\omega }{\rho},
\end{equation}
leading to the asymptotic estimate
\begin{eqnarray}
  \label{eq:asymest}
  i_k^{(n)}&\stackrel{k \to \infty}{=}& \left(\frac{k\omega}{\rho}\right)^{-\deltaj-n}\int_0^\infty
  dye^{-\rho y}y^{\omega
  k+n-s-1}\left[1+\mathcal{O}(1/k)\right]\nonumber
 \\ &=& \left(\frac{\omega
  k}{\rho}\right)^{-\deltaj-n}\rho^{-k\omega -n+s}\Gamma(k\omega +n-s)\left[1+\mathcal{O}(1/k)\right].
\end{eqnarray}
The behavior of the strong-coupling coefficients $b_k^{(n)}$ for large $k$
is obtained with the help of the identity
\begin{equation}
  \Gamma(z)\Gamma(1-z)=\frac{\pi}{\sin \pi z}
\end{equation}
and Stirling's formula (\ref{gammawachstum}), yielding
\begin{eqnarray}
  \label{eq:asymalpha}
 \!\!\!\!\!\!\!\!\!\!\! b_k^{(n)} &\stackrel{k \to \infty}{=}&  \gamma  \sin
  \pi[k(\omega-1)+\betaj +(1/\omega-1)s]
\left[-\frac{(1-\omega)^{(1-\omega)}}{(\sigmaj\rho)^\omega}\right]^k
  k^{\gamma_1}[1+\mathcal{O}(1/k)] .
\end{eqnarray}
The values of the real constants $\gamma ,\,\gamma_1$ will not be needed
in the upcoming discussions, and
are therefore not calculated explicitly.

Equation (\ref{eq:asymalpha}) shows
that the strong-coupling expansion (\ref{stark}) has a convergence radius
\begin{equation}
\label{eq:singularity}
  |g_B|\ge\frac{(\rho\sigmaj)^\omega}{(1-\omega)^{1-\omega}},
\end{equation}
which means that the basis functions $I_n(g_B)$, and certainly also $f(g_B)$
itself,
 possess
additional singularities beside $g_B=0$.
The parameter $ \rho $ will be optimally adjusted to match the positions of
these singularities.

\subsection{Taylor Series of Basis Functions}
For reexpanding $f_L(g_B)$ in terms of the basis functions
$I_n(g_B)$, we must know their Taylor series.
These are obtained by
substituting into (\ref{intdarst}) the variable $ y$  by $g_B y'$, and
expanding the integrand of (\ref{intdarst}) in powers of $g_B$.
After performing the integrals over $y'$ and $t$,
we find
\begin{equation}
\label{asymin}
  I_n(g_B)=
  \sum_{k=n}^\infty f^{(n)}_k g_B^k,
\end{equation}
with the coefficients
\begin{equation}
\label{asymink}
  f^{(n)}_k =
  \frac{1}{\omega}
  \frac{\Gamma(\betaj )\Gamma(k/\omega-s/\omega)}
  {\Gamma(k(1/\omega-1)+\betaj )} \sum_{j=0}^{k-n}{-\deltaj-n \choose j}\frac{(-\rho)^{k-n-j}}{(k-n-j)!}\sigmaj^k.
\end{equation}
The coefficients in the last sum arise from the $t$-integral:
\begin{eqnarray}
\label{eq:asymsum}
  \!\!\!\!\!\!\!\!\!\sum_{j=0}^{k-n}{-n-\deltaj \choose
    j}\frac{(-\rho)^{k-n-j}}{(k-n-j)!}&=&\frac{(-1)^{k-n}}{\Gamma(k-n+1)\Gamma(n+\deltaj)}\int_0^\infty dt e^{-t}t^{\deltaj+n-1}(\rho+t)^{k-n} .
\end{eqnarray}
For large $k$,
 the integral
may be evaluated with the help of the
saddle-point approximation. Using this and  Stirling's formula
(\ref{gammawachstum}), we find
\begin{eqnarray}
  \label{eq:asymsum}
  \!\!\!\!\!\!\!\!\!\sum_{j=0}^{k-n}{-n-\deltaj \choose
    j}\frac{(-\rho)^{k-n-j}}{(k-n-j)!}
&\stackrel{k \to \infty}{=}&
\frac{(-1)^{k-n}e^\rho}{\Gamma(\deltaj+n)}k^{\deltaj+n-1}
\left[1+\mathcal{O}(1/k)\right].
\end{eqnarray}
Inserting this into (\ref{asymink}) and using
once more Stirling's formula, we obtain
for the expansion coefficients $f_k^{(n)}$
the following factorial growth
\begin{eqnarray}
\label{asymwachs}
  f_k^{(n)} &\stackrel{k \to \infty}{=}& \frac{(-1)^n e^\rho\Gamma(\betaj )}{\sqrt{2\pi}\Gamma(\deltaj+n)}
  (1-\omega)^{1/2-\betaj }\omega^{\betaj -1+s/\omega}
  k^{ \deltaj-\betaj+n-3/2 -s/\omega}
  \left[-\frac{\sigmaj}{\omega(1-\omega)^{1/\omega-1}}\right]^kk! \nonumber
\\ &&\hspace{9cm}\times [1+\mathcal{O}(1/k)].
\end{eqnarray}

For an optimal reexpansion
(\ref{complete lis}),
we shall choose the free parameters
of the basis functions   $I_n(g_B,\omega,s,\rho,\sigmaj,\deltaj,\betaj )$
to match the large-order behavior of the coefficients $f_k$ in
(\ref{eq:wachs}).

\subsection{Convergence Properties of Resummed Series}

We shall now discuss the
speed of convergence of the resummation procedure.
For this it will be sufficient
to estimate the convergence
of the strong-coupling coefficients $b_k^L$
of the
approximations $f_L(g_B)$
 against the
true
strong-coupling coefficients $b_k$
in (\ref{stark}).  The convergence for arbitrary values of
$g_B$ will always be better than that.
Such an estimate is possible
by looking
at
the large-$n$
behavior of the expansion coefficients $b_k^{(n)}$
in the strong-coupling expansion
of $I_n(g_B)$.
in (\ref{instark}).
This is determined by the saddle point approximation to the
integral $i_k^{(n)}$ in Eq.~(\ref{koeffstarkin}), which we rewrite as
\begin{equation}
i_k^{(n)}=\int _0^\infty dye^{-\rho  y -n\ln\left(1+1/y\right)}(1+y)^{-\deltaj}y^{k\omega -s-1}.
\end{equation}
The saddle point lies at
\begin{equation}
    y_s=\sqrt{\frac{n}{\rho}}\left[1+\mathcal{O}(1/\sqrt{n})\right].
\end{equation}
At this point, the total exponent in the integrand is
\begin{equation}
  - \rho y_s-n\ln\left(1+\frac{1}{y_s}\right)= -2\sqrt{ \rho  n  }
  \left[1+\mathcal{O}(1/\sqrt{n})\right],
\end{equation}
implying the large-$n$ behavior
\begin{equation}
  b_k^{(n)}\stackrel{n \to \infty}{=}\mathrm{const.}\times n^{k\omega
  -s-1-\deltaj} e^{-2\sqrt{ \rho  n}}
  \left[1+\mathcal{O}(1/\sqrt{n})\right].
\label{@is implies}
\end{equation}

The strong-coupling coefficients $b_k^L$
of the approximations $f^a_L(g_B)$
are
 linear combinations of the coefficients $b_k^{(n)}$ of the basis
functions $I_n(g_B)$:
\begin{equation}
\label{eq:starkkoeff}
b_k^{L}=\sum_{n=0}^L b_k^{(n)}h_n.
\end{equation}
The speed of convergence
with which the $ b_k^{L}$'s
approach
 $ b_k$
 as the number $L$ goes to infinity
 is governed by the growth with $n$
of the reexpansion coefficients
$h_n$ and of the coefficients $b_k^{(n)}$ in Eq.~(\ref{@is implies}).
We shall see that for the series to be resummed, the
 reexpansion coefficients $h_n$ will grow at most like some power $n^r$,
implying that the approximations $b_k^L$ approach their $L\rightarrow \infty$
-limit $b_k$
with an error proportional to
\begin{equation}
b_k^{L} -b_k\sim L^{r+k\omega -s-\deltaj-1/2}\times e^{-2\sqrt{ \rho  L}}.
\label{@convergenceapp}\end{equation}
The leading exponential falloff of the error $ e^{-2\sqrt{ \rho  L}}$
 is independent of the other parameters
in the basis functions $I_n(g_B,\omega,p,\rho,\sigmaj,\deltaj,\betaj )$
which still need adjustment.
This is the important  advantage
of the present resummation method
with respect to variational perturbation theory \cite{PI,PI3}
where the error decreases merely
like $e^{-{\rm const}\times L^{1- \omega }}$ with $ 1-\omega$ close to $1/4$.

The nonexponential prefactor in Eq.~(\ref{@convergenceapp}) depends
on  the  parameters
in $I_n(g_B,\omega,p,\rho,\sigmaj,\deltaj,\betaj )$.
Some of them are related to observables, others are
free and may be chosen to optimize the convergence.

\subsubsection{Parameters $s$ and $\omega$}
The perturbation expansions
for the critical exponents are  power series in the  bare coupling
constant $g_B$ whose strong-coupling limit is a constant \cite{PI3,seven}.
The same is true for the series expressing the renormalized coupling constant
$g$
in powers of the bare coupling constant.
This implies that the
growth parameter $s$ for the basis functions
$I_n(g_B)$
is equal to zero in all cases.
The constant asymptotic values
are approached with the subleading powers $ 1/g_B^{ k\omega -s}$,
where  $\omega$
is a universal experimentally measurable
critical exponent.

\subsubsection{Parameter $ \sigma $}
In the ordinary Borel-transformation,
the parameter $\sigmaOj$ in the
large-order behavior
of the expansion coefficients $f_k$ in
Eq.~(\ref{eq:wachs}),
which is determined directly by the
inverse value of the reduced action of the classical solution to the field
equations,
specifies also the position of the
singularity on the negative $t$ axis in $B(t)$.
In our  transform $\tilde{B}(y)$,
the singularity
position of the singularity
is proportional to $ \alpha $, with an
$  \omega $-dependent prefactor. It lies at
[see  Eq.~(\ref{posbor})]
\begin{equation}
  \label{connection}
 \sigmaj=\alpha\omega(1-\omega)^{1/\omega-1}.
\label{@thischoice}\end{equation}
This value of $\sigma$ ensures that
the expansion coefficients $f_k^n$
of the basis functions $I_n(g_B)$ in Eq.~(\ref{asymwachs})
grow for large $k$
with the same factor $(-\alpha )^k$
as the expansion coefficients
for $f(g_B)$ in Eq.~(\ref{eq:wachs}).

The conformal mapping (\ref{eq:konfaby})
 maps the singularity at $t=-1/\sigma$ to $ \kappa =\infty$,
and converts the cut along
the negative into a cut in the $\kappa$-plane
from $1$ to $\infty$.
The growth
of the reexpansion coefficients $h_n$
with $n$ is therefore
determined by the nature of the singularity of $\tilde{B}(y)$ at $\infty$.

In the upcoming applications to critical exponents
it will turn out that the value (\ref{@thischoice})
following from
the inverse action of
the solution to the classical field equations
and $ \omega $ will not yield the
 fastest convergence of the approximations $f^L(g_B)$ towards $f(g_B)$, but
 that a slightly smaller value gives better results.
 This seems to be due to the fact that the classical solution
gives only the nearest singularity in the  hyper-Borel transform
$\tilde B(y)$ of $f(g_B)$. In reality, there are many additional cuts
from other fluctuating field configurations
which determine the size of
the expansion coefficients $f_k$ at pre-asymptotic orders $k$.
Since the  few known $f_k$`s  are always pre-asymptotic,
they are best
accounted for by an effective  shift of the position of the singularity
into the direction of the additional cuts at larger negative $y$,
corresponding to a smaller $ \sigma $.

\subsubsection{Parameter $\rho$}
According to Eq.~(\ref{eq:singularity}),
the parameter $ \rho $
determines the radius
of convergence of the strong-coupling expansion of the basis functions
$I_n(g_B)$.
It should therefore  be adjusted to fit optimally
the corresponding radius
of the original function $f(g_B)$.
Since we do not know this radius, this adjustment will be done
phenomenologically by varying $ \rho $
to optimize the speed of convergence.
Specifically,
we shall  search at each order  $L$ for a vanishing highest  reexpansion
coefficient $h_L$ or, if it does not vanish anywhere, for a vanishing
derivative with respect to $\rho$:
\begin{equation}
  h_L(\rho)=0,\;\;\mathrm{or}\;\;\frac{dh_L(\rho)}{d\rho}=0.
\label{@hLrho}\end{equation}

\subsubsection{Parameter $\delta$}
From Eq.~(\ref{asymwachs})] we see that
the parameter $ \delta $ influences the power $k^ \beta $
in the large-order behavior (\ref{eq:wachs}).
By comparing the two equations,
we identify the growth parameter $ \beta $ of $I_n(g_B)$
as being
\begin{equation}
  \beta =\delta - \betaj-3/2-s/\omega+n.
\label{@have set}
\end{equation}

At first it appears to be impossible
to give {\em all\/}
basis functions $I_n(g_B)$
the same growth power $\beta $  in
(\ref{asymwachs}) by simply letting $ \delta $
depend on the order $n$
as required by (\ref{@have set}).
If we were to do this, we would have to assign to $\delta$ the value
\begin{equation}
\delta = \delta _n\equiv \beta + \betaj+3/2+s/\omega-n,
\label{@have setdel}
\end{equation}
which depends on the index $n$ of the function $I_n(g_B)$,
and this means that we perform an analytical continuation of
the  power series expansion of $\tilde{B}(y)$ by reexpanding
it as follows:
\begin{equation}
  \label{eq:worsean}
  \tilde B(y)=\sum_{k=0}^\infty \tilde B_k
  y^k=e^{-\rho\sigma y}(1+\sigma y)^{-\delta}\sum_{k=0}^\infty
  h_k(\sigma y)^k.
\end{equation}
But the series in this formula which is obtained from the series of
$\tilde{B}(y)$ by removing a simple factor still has the same finite radius
of convergence and could not be used to estimate $\tilde{B}(y)$ for
large values of $y$ needed to perform the back-transform (\ref{umkehrtrans}).
It is, however,  possible to sidetrack this problem
by letting  $\rho$ grow linearly with the
order $L$. Then the exponential factor of
(\ref{eq:worsean}) suppresses the integrals over $y$ for large $y$
sufficiently to make the  divergence of the reexpanded series
(\ref{eq:worsean}) at large $y$
irrelevant.
If we determine $\rho$ from the condition (\ref{@hLrho}),
the growth of $ \rho $ with
$L$ turns out to emerge by itself.

\subsubsection{Parameter $\betaj $}
The parameter $\betaj$ has two effects.
From Eq.~(\ref{koeffstark}) we see that for
\begin{equation}
k>k_c\equiv\frac{\betaj +(1/\omega-1)s}{1- \omega }
\label{@}\end{equation}
 the signs of the strong-coupling expansion coefficients
start to alternate irregularly.
This irregularity weakens the convergence of the
higher strong-coupling coefficients $b_k^L$ with $k>k_c$ against
$b_k$. The convergence can therefore be improved
by choosing a
$\betaj$ which grows proportionally to the order $L$ of the approximation.

In addition, $ \betaj  $ appears in
 the power of $k$ in (\ref{asymwachs}), which is a consequence of the fact that
it determines the nature of the cut
in $\tilde B(y)$ in the complex $y$-plane starting at $y=-1/ \sigma $
[see Eq.~(\ref{eq:analytfort})].

If we expand both sides of (\ref{eq:analytfort}) in powers of
$\kappa=\sigmaj y/(1+\sigmaj y)$
and compare the coefficients of  powers of $\kappa$, it is easy to
write down an
explicit formula for the reexpansion coefficients $h_n$
in terms of the coefficients $\tilde B_j$ of $\tilde B(y)$
by
\begin{equation}
  \label{eq:reexpcoeff}
 h_n=\sum_{k=0}^n\sum_{j=0}^k
  \frac{\tilde B_j\sigmaj^{-j}\rho^{k-j}}{(k-j)!}{\deltaj+n-1\choose n-k},
\end{equation}
where $\tilde B_j$
 are obtained from the original expansion coefficients
$f_k$ of $f(g_B)$ by relation
(\ref{modbor}).

Before beginning with
the resummation of the perturbation expansions
for the critical exponents of $\phi^4$-field theories,
it will be useful
to obtain a feeling for the quality
of the above-developed resummation procedure,
in particular  for the significance of the
parameters upon the speed of convergence. We do this
by resumming the often-used example of an asymptotic series,
the perturbation expansion
of the ground state energy of the anharmonic oscillator.

\subsection{Resummation of Ground State Energy of Anharmonic Oscillator}
Consider the one-dimensional anharmonic oscillator with the Hamiltonian
\begin{equation}
  \label{eq:hamosz}
  H=\frac{p^2}{2}+m^2 \frac{x^2}{2}+g_Bx^4.
\end{equation}
In this quantum mechanical system,
there is no need to distinguish bare
and renormalized coupling constants,
but since the previous resummation formulas were all
formulated in terms of $g_B$
we shall keep this notation also here.
The ground state energy has a perturbation expansion
\begin{equation}
E^{(0)}(g_B)= \sum _k^\infty f_k g_B^k,
\label{@pertexpanE}\end{equation}
whose coefficients can be calculated via the Bender-Wu recursion relation
\cite{bender}
to arbitrarily high orders, with a large-order behavior
\begin{equation}
\label{largeosz}
f_k = -\sqrt{\frac{6}{\pi^3}}\;k!(-3)^kk^{-1/2} [1+\mathcal{O}(1/k)].
\end{equation}
By comparison with (\ref{eq:wachs})
we identify the growth parameters
\begin{equation}
  \alpha =3,~~~~\beta =-1/2.
\label{@idenbeta}\end{equation}
A scale transformation $x \to g^{1/6}x$
applied to the Hamiltonian (\ref{eq:hamosz}) reveals the
scaling property  \cite{simon}
for the energy as a function of $g_B$ and $m^2$:
\begin{equation}
  \label{eq:skaling}
  E(m^2,g_B)=g_B^{1/3}E(g_B^{-2/3}m^2,1)  .
\end{equation}
Combining this with the knowledge \cite{simon} that $E(m^2,1)$ is an analytic
function
at $m^2=0$, we see that $E(1,g_B)$ possesses a power series  expansion of the form
(\ref{stark}), with the parameters
\begin{equation}
s=1/3,~~~~\omega=2/3.
\label{@sandom}\end{equation}
Inserting the latter number together  with $ \alpha $
from Eq.~(\ref{@idenbeta}) into (\ref{@thischoice}), we
identify
\begin{equation}
 \sigma =\frac{2}{ \sqrt{3} }.
\label{@}\end{equation}
The ground state energy $E^{(0)}(g_B)$ obeys a once-subtracted dispersion relation \cite{simon}:
\begin{equation}
E^{(0)}(g_B)=\frac{1}{2}+\frac{g_B}{\pi}\int_0^\infty
\frac{dg'_B}{g'_B}\frac{\Im E^{(0)}(-g'_B)}{g'_B+g_B}.
\end{equation}
The perturbation expansion
(\ref{@pertexpanE}) is obtained from this by
expanding $1/(g'_B+g_B)$ in powers of
$g_B$, and performing the integral term by term.
This shows explicitly that
the large-order behavior (\ref{largeosz}) is caused by
 an
imaginary part
\begin{equation}
  \label{eq:imosz}
  \Im E^{(0)}(-|g_B|){=}\sqrt{\frac{6}{\pi}}\sqrt{\frac{1}{3|g_B|}}e^{-1/3|g_B|}[1+\mathcal{O}(|g_B|)]
\end{equation}
near the tip of the left-hand cut in the complex $g_B$-plane,
in agreement with the general form
(\ref{eq:imp})
associated with the large-order behavior
(\ref{eq:wachs}).

Let us now specify the parameter $ \delta $.
We shall do this here in an $ n $-dependent way
using
Eq.~(\ref{@have setdel}), which now reads with
(\ref{@sandom}):
\begin{equation}
 \delta = \delta _n\equiv  \beta _0+3/2-n.
\label{@deltanch}\end{equation}
The corresponding
 basis functions
\begin{equation}
  \label{eq:iosz}
  I_n(g_B,2/3,1/3,\rho,2/\sqrt{3},\betaj +3/2-n,\betaj ),
\end{equation}
  have then all the
same large-order growth parameter $ \beta $ in (\ref{eq:wachs}).

The two parameters $\rho$  and $ \betaj$ are still arbitrary.
The first is determined by an order-dependent optimization
of the approximations
via the conditions
(\ref{@hLrho}). The  best choice of $\betaj$ will be
made differently depending on the regions
of $g_B$.

Let us test the convergence of our algorithm at small negative coupling
constants
$g_B$,
i.e., near the tip of the left-hand cut in the complex $g_B$-plane.
We do this by  calculating the
prefactor $\gamma $ in the large-order behavior (\ref{eq:wachs}).
In this case the convergence turns out to be fastest
by giving the parameter
$\betaj$ a small value, i.e. $\betaj=2$.
With the large-order behavior
(\ref{asymwachs}) of the basis functions $I_n(g_B)$, we find
the resummed functions $f_L(g_B)$ of $L$th order
$\sum _{n=0}^L h_n I_n(g_B)$ to have
a large-order behavior
 (\ref{eq:wachs})  with a prefactor
\begin{equation}
  \label{eq:calcco}
  \gamma_{L}=\frac{e^\rho
  \Gamma(\betaj )}{\sqrt{2\pi}\Gamma(\deltaj)}\sum_{k=0}^L (-1)^kh_k.
\end{equation}

\begin{figure}[b]
\setlength{\unitlength}{1cm}
\begin{picture}(7,4.2)
\put(0,0){\IncludeEpsImg{48.5mm}{48.5mm}{0.7000}{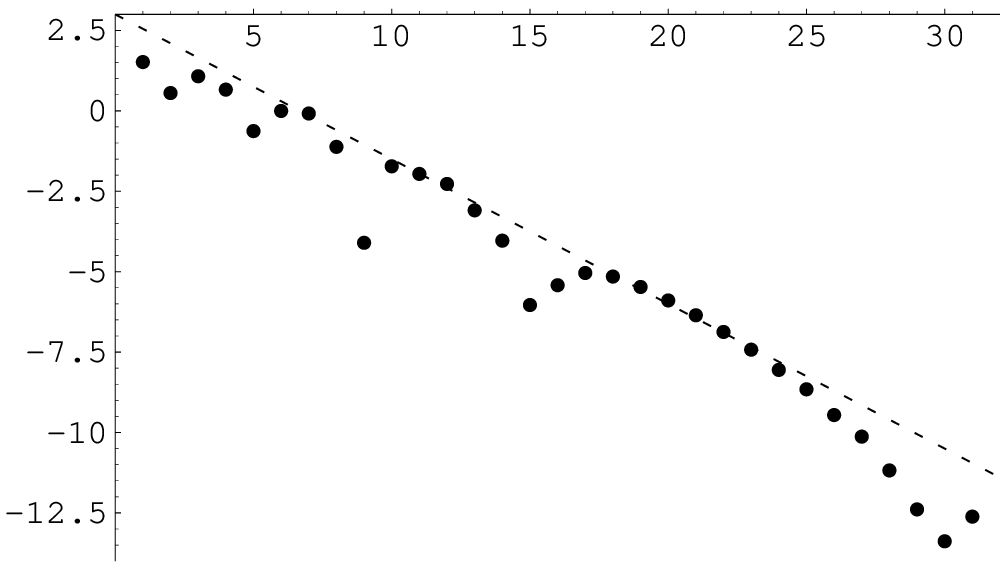}}
\put(6.85,4){\fsz$L$}
\put(3.4,2.8){\fsz$\ln|\gamma_L- \gamma| \approx 3-0.45L$}
\end{picture}
~~~~            ~~~~~
\raisebox{-3mm}{\begin{picture}(7,4.2)
\put(0,0){\IncludeEpsImg{48.5mm}{48.5mm}{0.7000}{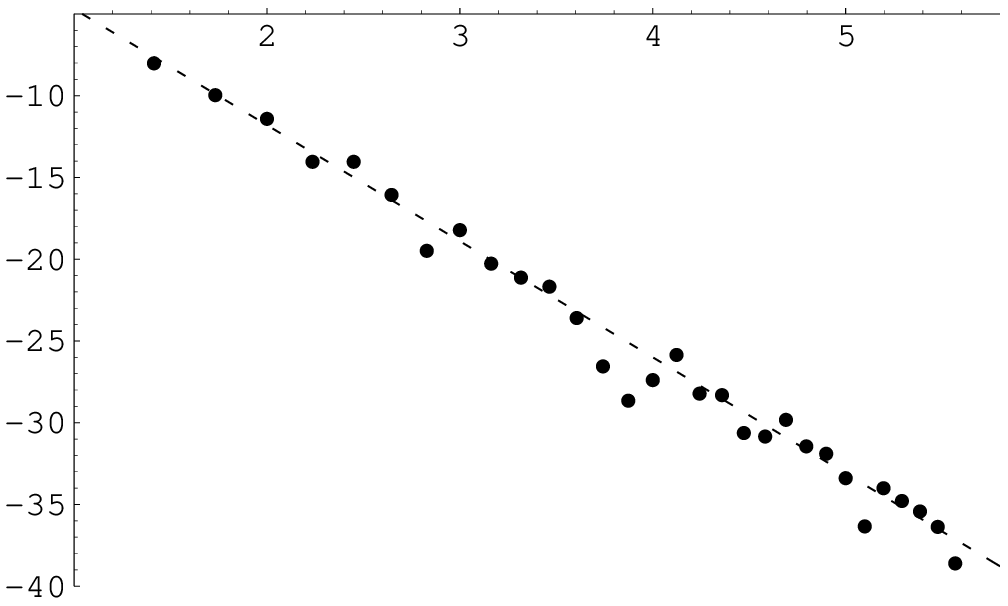}}
\put(6.5,4.3){\fsz$\sqrt{L}$}
\put(3.4,2.8){\fsz$\ln|b_0 ^L-b_0 | \approx 2.4-7.1\sqrt{L}$}
\end{picture}}
\caption{Logarithmic plot of the convergence behavior of the
successive approximations to the
  prefactor  $\gamma ^L$ in the large-order
behavior (\protect\ref{@normlfacgam}),
 and of
the leading strong coupling coefficient $b_0^L$.}
\label{imkonv}
\end{figure}

The values of these sums for increasing ${L}$
are shown in Fig.~\ref{imkonv}.
They converge exponentially fast against the
exact limiting value
\begin{equation}
 \gamma =\sqrt{\frac{6}{\pi^3}},
\label{@}\end{equation}
with superimposed oscillations.
The oscillations are of the same kind as
those observed in  variational perturbation theory
for the convergence of the approximations to the
strong-coupling coefficients $b_k$ (see Figs.~5.19 and 5.20 in Ref.~\cite{PI})
Also here, the strong-coupling coefficients $b_k^L$
converge exponentially fast
towards $b_k$, but with a larger power of $L$ in the exponent
of the last term $\approx e^{-{\rm const} \times  \sqrt{L}}$
[see Eq.~(\ref{@convergenceapp})], rather
than $\approx e^{-{\rm const}\times   L^{1/3}}$
for variational perturbation theory [see Eq.~(5.199) in Ref.~\cite{PI}].
This is seen on the right-hand side of Fig.~\ref{imkonv}.

We have applied our resummation method
to the first 10 strong-coupling coefficients
using the expansion coefficients
$f_k$ up to order $70$.
The results are
shown
in Table~1.
\begin{table}[tbh]
\makebox[\textwidth]{\begin{math}
\begin{array}{|r|r@{.}l|} \hline \rule{0pt}{2.3ex}
 n & \multicolumn{2}{c|}{b_n}
\\[0.5ex] \hline \rule{0pt}{2.3ex}
 0 &  0&667\,986\,259\,155\,777\,108\,270\,962\,02  \\
 1 &  0&143\,668\,783\,380\,864\,910\,020\,319      \\
 2 & -0&008\,627\,565\,680\,802\,279\,127\,963      \\
 3 &  0&000\,818\,208\,905\,756\,349\,542\,41       \\
 4 & -0&000\,082\,429\,217\,130\,077\,219\,91       \\
 5 &  0&000\,008\,069\,494\,235\,040\,964\,75       \\
 6 & -0&000\,000\,727\,977\,005\,945\,772\,63       \\
 7 &  0&000\,000\,056\,145\,997\,222\,351\,17       \\
 8 & -0&000\,000\,002\,949\,562\,732\,709\,36       \\
 9 & -0&000\,000\,000\,064\,215\,331\,956\,97       \\
 10&  0&000\,000\,000\,048\,214\,263\,789\,07       \\[0.5ex] \hline
\end{array}
\end{math}}
\label{strongcoeff}
\caption{Strong-coupling coefficients $b_n$ of the $70$-th order approximants
  $E_{70}^{0}(g)=\sum_{n=0}^{70} h_nI_n(g)$ to the ground state energy
  $E^{0}(g)$ of the anharmonic oscillator. They have the same accuracy as the variational
perturbation-theoretic calculations up to order $251$ in
  Refs.~\protect\cite{JK,PI}.}
\end{table}
Comparison with a similar table
in Refs.~\cite{JK,PI}
shows that the new resummation method
yields in $70$th order the same accuracy as
variational perturbation theory did in $251$st order.
In all cases the optimal parameter $\rho$ turns out to be a slowly
growing function with $L$.

In the strong-coupling regime,
the convergence is fastest
by choosing for $ \betaj  $
an $L$-dependent value
\begin{equation}
  \label{eq:scalebeta}
  \betaj =L.
\end{equation}
Note that this choice of $\betaj $ ruins the convergence to the imaginary part
for small negative $g_B$ which was resummed
best with $ \beta _0=2$.

\subsection{Resummation of Critical Exponents}
Having convinced ourselves
of the fast convergence of our new resummation method,
let us now turn to the
perturbation  expansions of the
O($N$)-symmetric $\phi^4$-theories in powers of the bare coupling constant
$\bar{g}_B$, defined by the euclidean
action
\begin{equation}
  \label{eq:eucact}
  \mathcal{A}=\int
  d^Dx \left\{\frac{1}{2}[\partial\phi_0(x)]^2+\frac{1}{2}m_0^2
  \phi_0^2(x)+2\pi {\bar g}_B[\phi_0^2(x)]^2\right\}
\end{equation}
in $D=3$ dimensions. The field $\phi_0$
is an $N$-component vector
$\phi_0=(\phi_0^1,\phi_0^2,\ldots,\phi_0^N)$,
and the action is
O($N$)-symmetric.
We define renormalized mass $m$ and field strength
by parametrizing the behavior of the connected two point function $G^{(2)}$
 in momentum
space near zero momentum as
\begin{equation}
  \label{eq:defrenconst}
  G^{(2)}(p,\alpha;-p,\beta) = Z_\phi\frac{\delta_{\alpha
  \beta}}{m^2+p^2+\mathcal{O}(p^4)}.
\end{equation}
The renormalized coupling constant $g$ is defined by
the value of the connected four-point function
at zero momenta:
\begin{equation}
  G^{(4)}(0,\alpha;0,\beta;0,\gamma;0,\delta) =
  m^{-4-D}Z_\phi^2\,g(\delta_{\alpha\beta}\delta_{\gamma\delta} +
  \delta_{\alpha\gamma}\delta_{\beta\delta} +
  \delta_{\alpha\delta}\delta_{\beta\gamma}).
\end{equation}
If we introduce the dimensionless bare coupling constant $g_B \equiv
\bar{g}_B/m$, the critical exponents are defined by
\begin{eqnarray}
  \eta(g_B)=g_B\frac{d}{dg_B}\log Z_\phi,\nonumber \\
  2-\nu(g_B)^{-1}=g_B\frac{d}{dg_B}\log\frac{m_0^2}{m^2}.
\end{eqnarray}
The following expansions for the critical indices in the bare dimensionless
coupling constant are
available \cite{sokolov} in the literature for all O($N$):
\begin{eqnarray}
\eta(g_B) \!\!\!\!\!\!\!
&&=(16/27+8N/27) g_B^2+(-9.086537459-5.679085912N-0.5679085912N^2)g_B^3 \nonumber\\
&&+ (127.4916153+94.77320534N+17.1347755N^2+0.8105383221N^3) g_B^4 \nonumber\\
&&+ (-1843.49199-1576.46676N-395.2678358N^2-36.00660242N^3 \nonumber\\
&& \hspace{1cm} -1.026437849N^4) g_B^5 \nonumber\\
&&+ (28108.60398+26995.87962N+8461.481806N^2+1116.246863N^3 \nonumber\\
&& \hspace{1cm}+62.8879068N^4+1.218861532N^5)g_B^6\;,  \\
2-\nu^{-1}(g_B)  \!\!\!\!\!\!\!
&&=g_B(2+N)+(523/27+316N/27+N^2) g_B^2 \nonumber \\
&&+ (229.3744544+162.8474234N+26.08009809N^2+N^3) g_B^3 \nonumber \\
&&+ (-3090996037-2520.848751N-572.3282893N^2-44.32646141N^3 -N^4)g_B^4\nonumber\\
&&+ (45970.71839+42170.32707N+12152.70675N^2+1408.064008N^3 \nonumber\\
&& \hspace{1cm} +65.97630108N^4+N^5)g_B^5 \nonumber \\
&&+ (-740843.1985-751333.064N-258945.0037N^2-39575.57037N^3 \nonumber\\
&& \hspace{1cm} -2842.8966N^4-90.7145582N^5-N^6)g_B^6\;.
\end{eqnarray}
In addition, seventh order coefficients have been calculated
for $N=0,1,2,3$:
\cite{nickel}:
\begin{equation}
  \eta^{(7)}=\left \{\begin{array}{c}
   -45387.48927
\\ -114574.4876
\\ -241424.7646
\\ -454761.4731
\end{array} \right \}g_B^7,\;\;
\nu^{-1\;(7)}=\left \{\begin{array}{c}
   -12792269.773
\\ -33711416.972
\\ -73780809.849
\\ -143831857.01
\end{array} \right \}g_B^7\;\;
\mbox{for}\;\;
\left \{\begin{array}{c}
   N=0
\\ N=1
\\ N=2
\\ N=3
\end{array} \right \}.
\end{equation}
When approaching the critical point, the renormalized mass $m$ tends to zero,
so that the problem is to find the strong-coupling limit of these
expansions.
In order to have the critical exponents approach a constant value,
the power $s$ in Eq.~(\ref{stark}) must be set equal to zero.

\begin{figure}[h]
\unitlength=1mm
\begin{picture}(48.5,150)

\put(-2,104){\IncludeEpsImg{48.5mm}{48.5mm}{0.7000}{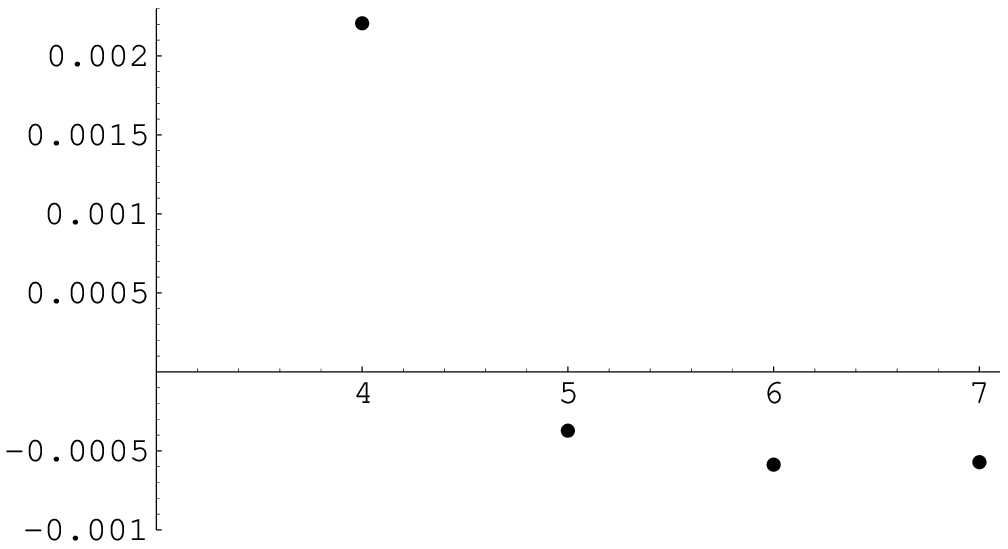}}

\put(88,104){\IncludeEpsImg{48.5mm}{48.5mm}{0.7000}{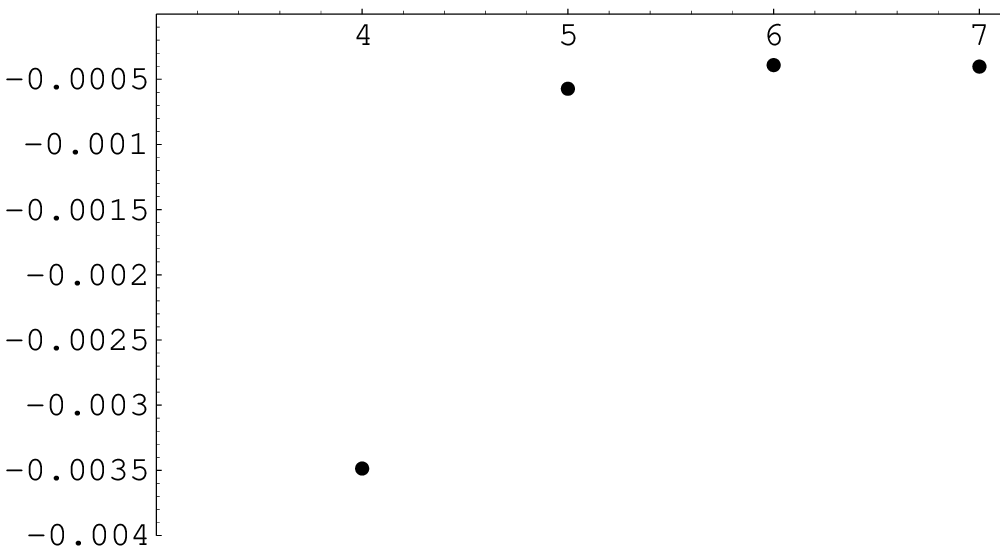}}
\put(-2,52){\IncludeEpsImg{48.5mm}{48.5mm}{0.7000}{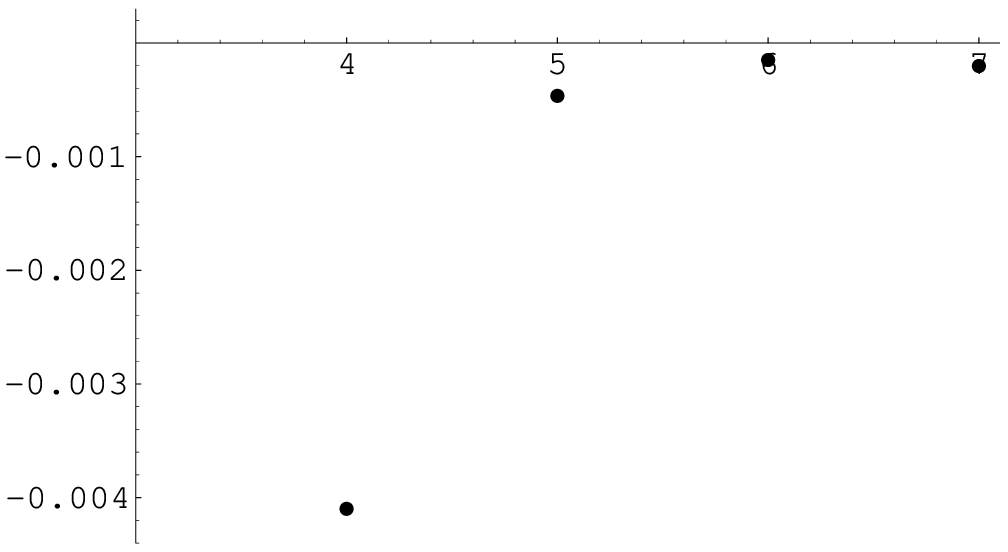}}
\put(88,52){\IncludeEpsImg{48.5mm}{48.5mm}{0.7000}{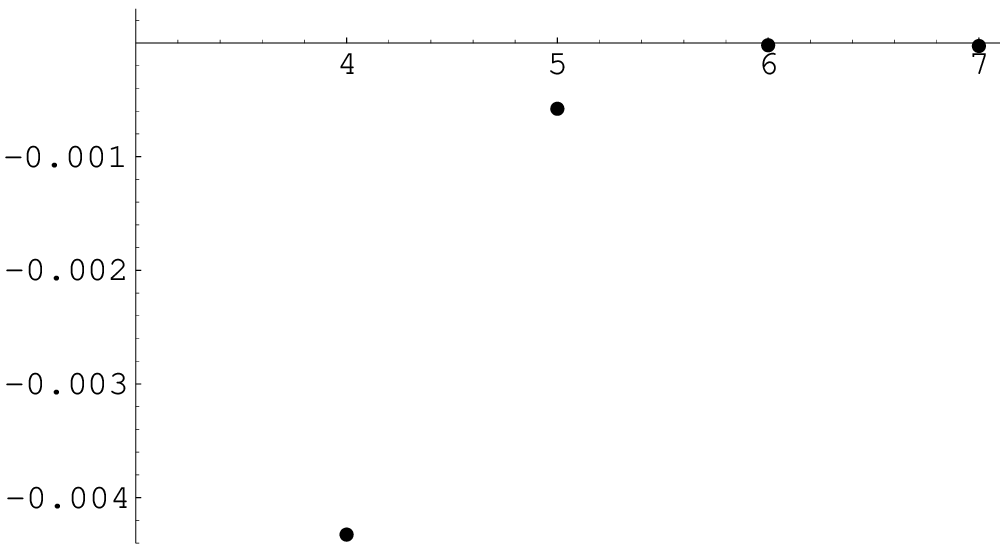}}
\put(-2,0){\IncludeEpsImg{48.5mm}{48.5mm}{0.7000}{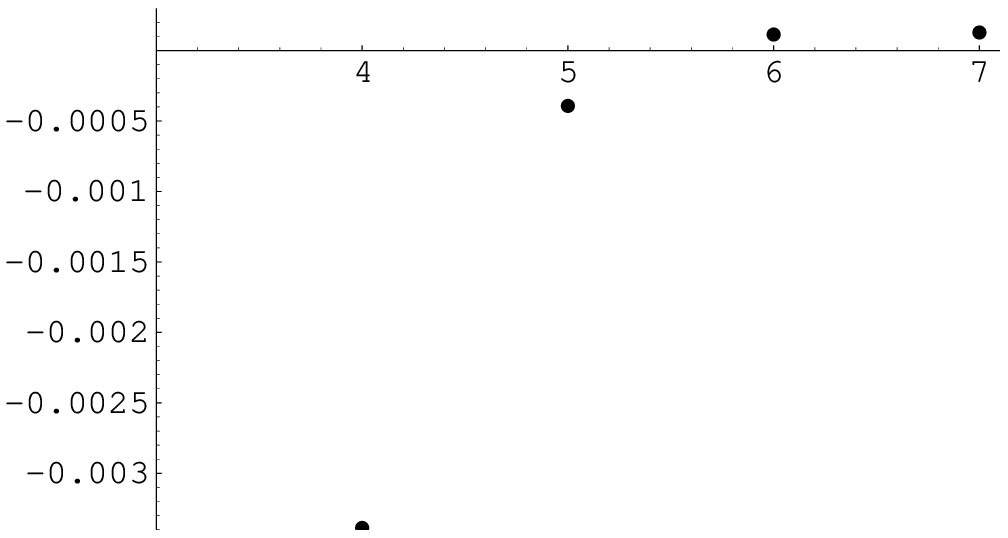}}

\put(35,129){\fsz$\omega=0.786$}
\put(11,111){\fsz$\beta^*_L$}
\put(35,142){\fsz$0.0003 \pm0.00015$}

\put(125,129){\fsz$\omega=0.789$}
\put(103,111){\fsz$\beta^*_L$}
\put(125,140){\fsz$0.00012 \pm0.00015$}

\put(35,77){\fsz$\omega=0.792$}
\put(11,59){\fsz$\beta^*_L$}
\put(35,88){\fsz$0 \pm0.00015$}

\put(125,77){\fsz$\omega=0.795$}
\put(103,59){\fsz$\beta^*_L$}
\put(125,86){\fsz$-0.00023 \pm0.00015$}

\put(35,23){\fsz$\omega=0.798$}
\put(11,7){\fsz$\beta^*_L$}
\put(35,33){\fsz$-0.00043 \pm0.00015$}
\end{picture}
\caption{Convergence of strong-coupling limits of the
  $\beta$-function (\ref{@betafunc}) for $N=1$ and different values of
 $\omega$. The upper and lower dashed lines denote the range of the
 $L \to \infty$ limit of $\beta^*_L$ from which the value of $\omega$ is
 deduced in Fig.~\ref{erom}.}
\label{fig1}
\end{figure}

In contrast to the quantum-mechanical discussion in the last section,
the exponent $ \omega $ governing the approach to the scaling limit
is now unknown, and must also be determined from the
available perturbation expansions.
As in  Ref.~\cite{PI3,seven}, we
solve this problem by using the fact
that the existence of a critical point
implies
the renormalized coupling constant $g$ in powers of $g_B$
to converge against a constant renormalized
coupling $g^*$ for $m\rightarrow 0$.
The expansion of $g(g_B)$ is known
up to six loops
\cite{sokolov} for all O($N$):
 \begin{eqnarray}
g(g_B)
 \!\!\!\!\!\!\!
&&=g_B+(-8-N) g_B^2+(2108/27+514/27 N+N^2) g_B^3\nonumber \\
&& + (-878.7937193-312.63444671 N-32.54841303 N^2-N^3) g_B^4 \nonumber \\
&&+ (11068.06183+5100.403285 N+786.3665699 N^2+48.21386744 N^3+N^4) g_B^5
  \nonumber \\
&&+ (-153102.85023-85611.91996 N - 17317.7025 N^2-1585.114189 N^3 \nonumber\\
&& \hspace{1cm}-65.82036203 N^4-N^5)g_B^6 \nonumber \\
&&+  (2297647.148+1495703.313 N+ 371103.0896 N^2+44914.04818 N^3 \nonumber\\
&& \hspace{1cm} +2797.291579N^4+85.21310501 N^5+N^6)g_B^7\;.
\label{@gvongB}\end{eqnarray}
The convergence
against a fixed
coupling $g^*$ occurs
only for
the correct
value of $ \omega $ in the
 resummation functions
$
I_n(g_B,\omega,s,\rho,\sigmaj,\deltaj,\betaj )$.
At different values,
 $g(g_B)$ has some strong-coupling
power behavior  $g_B^s$ with  $s\neq 0$.
We may therefore determine
$ \omega $ by forming from (\ref{@gvongB})
a series for the power $s$,
\begin{equation}
s=\frac{d\log g(g_B)}{d\log g_B}=\frac{g_B}{g} g'(g_B),
\label{@}\end{equation}
resumming this for various values of $ \omega $ in the basis functions,
and finding the critical exponent $ \omega $ from the zero of $s$.
Alternatively, since  $g(g_B)\rightarrow g^*$, we can just as well
resum the series for $-gs$, which coincides with the
$ \beta $-function
of renormalization group theory [not to be confused with the
growth parameter $ \beta $ in (\ref{eq:wachs})]
\begin{equation}
  \beta(g_B)\equiv -g_B\frac{dg(g_B)}{dg_B}.
\label{@betafunc}
\end{equation}
If we denote its
strong-coupling  limit by $ \beta ^*$,
\begin{equation}
 \beta ^*\equiv  \beta (g_B)|_{g_B\rightarrow \infty},
\label{@}\end{equation}
we resum the expansion for $ \beta (g_B)$ to form
the approximations
\begin{equation}
 \beta_L(g_B)=\sum_{n=0}^Lh_nI_n(g_B,\omega),
\end{equation}
and plot the strong-coupling limits
of the $L$th approximations
$ \beta ^*_L$
for various values of $\omega$. This is shown in  Fig.~\ref{fig1}.
 From these plots we extract the critical exponent $ \omega $
by finding the  $\omega$-value for which the approximations
$ \beta ^*_L$
 extrapolate best to zero for $L \to \infty$,
taking into account that
the convergence is exponentially fast with superimposed oscillations.
These $ \omega $-values are called $ \omega _L$.

For these resummations, we must  of course
specify the remaining parameters in the basis functions
$I_n(g_B,\omega,0,\rho,\sigmaj,\deltaj,\betaj )$.
This can, in principle,
proceed as in the case of the
anharmonic oscillator.
The parameter $ \alpha $
is determined from  the action of a
classical instanton solution $\phi_c(x)$ of the field equations,
 and has for all  expansions the $N$-independent value \cite{parisi2}
\begin{eqnarray}
\label{@growal}
   \alpha =\frac{32\pi}{I_4}= 1.32997 \;,
\end{eqnarray}
where  $I_p\equiv\int d^Dx[\phi_c(x)]^p$ are integrals over
powers of $\phi_c(x)$.

To determine the parameter
$\delta$,
we recall the remaining
growth parameters $ \beta $ and $ \gamma $ of the large-order behavior
(\ref{eq:wachs}) of the perturbative series
for the critical exponents.
The growth parameter $\beta$ is given by
 the number of zero modes in the fluctuation spectrum
around this classical solution:
\begin{equation}
\left \{\begin{array}{l}
 \beta _   \eta
\\ \beta _ {\nu^{-1}}
\\  \beta _\beta
\end{array} \right \}=
\left \{\begin{array}{c}
   3+N/2
\\ 4+N/2
\\ 4+N/2
\end{array} \right \}
\end{equation}
The prefactors $\gamma$ in (\ref{eq:wachs})
requires the
calculation of the fluctuation determinants
around the classical solution, which yields in the case of
the $\beta$-function
\begin{equation}
  \gamma_\beta=\frac{2^{N/2+2}3^{-3/2}\pi^{-2}}{\Gamma(N/2+2)}\left(\frac{I_1^2}{I_4}\right)^2\left(\frac{I_6}{I_4}-1\right)^{3/2}D_L^{-1/2}D_T^{-(N-1)/2}\;.
\label{@normlfacgam}\end{equation}
where
$D_L$ and $D_T$ are characteristic
quantities of the longitudinal and transverse parts of the fluctuation
determinant, respectively. Their numerical values are  \cite{parisi2}
\begin{equation}
\label{const}
\begin{array}{|c c c c c c|} \hline\hline
\rule{0pt}{2.3ex}
  D_L & D_T & I_1 & I_4 & I_6 & H_3
\\\hline
%\\[-2pt]
%\rule{0pt}{2.3ex}
10.544\pm0.004 & 1.4571\pm0.0001 & 31.691522 & 75.589005 & 659.868352 &
13.563312 \\[2pt]
\hline\hline
\end{array}
\end{equation}
The parameters
$ \gamma _ \eta ,~ \gamma  _ {\nu^{-1}}~ $
are obtained from  $ \gamma _ \beta $ by:
\begin{equation}
\gamma_\eta=\gamma_\beta\frac{2H_3}{I_1D(4-D)}, \;\;\; \gamma_{\nu^{-1}} =
\gamma_\beta\frac{N+2}{N+8}(D-1)4\pi\frac{I_2}{I_1^2},
\label{@}\end{equation}
where $I_2=(1-D/4)I_4$ and $H_3$ is listed in (\ref{const}).
Note that the expansions
in powers of the renormalized coupling constant $g$
have the same parameters $ \alpha $ and
$\beta$, but different parameters
$ \gamma_R$.
These differ from the above $ \alpha $'s by a common factor:
\begin{equation}
\gamma_R=\gamma \; e^{-({N+8})/{\alpha}}.
\label{@}\end{equation}

From Eq. (\ref{@growal}), the parameter $ \sigma $ is
found using relation
(\ref{connection}).
It turns out, however, that
this value does not lead to an optimal convergence.
This can be understood
qualitatively
by observing that the large-order behavior
of the expansion coefficients of the critical exponents and of the $ \beta $-function
in powers of the bare coupling constant $g_B$ is not nearly as
precocious in reaching the large-order form
(\ref{eq:wachs})
as the corresponding  expansions
in powers of the renormalized
coupling constant $g$ (see Fig.~1 in Ref.~\cite{seven}).
The lack of precocity
here is
 illustrated
for the expansion coefficients $\beta_k$ of the $\beta$-function in
Table~2,
which gives the ratios of $\beta_k$ and their leading
asymptotic estimates $ \beta ^{\rm as}_k$:
\begin{table}[htb]
\makebox[\textwidth]{\begin{math}
\begin{array}{|c|c|c|c|c|} \hline
\rule{0pt}{2.3ex}
 N & 0 & 1 & 2 & 3 \\[2pt]\hline\hline
k & {\beta}_k/\beta ^{\rm as}_k &{\beta}_k/\beta ^{\rm as}_k &{\beta}_k/\beta ^{\rm as}_k & {\beta}_k/\beta ^{\rm as}_k \\[2pt]
  \hline
\rule{0pt}{2.3ex}
2  &0.57 &0.45 &0.35 &0.27  \\
3  &0.61 &0.45 &0.32 &0.22  \\
4  &0.73 &0.51 &0.34 &0.22  \\
5  &0.89 &0.61 &0.40 &0.25  \\
6  &1.07 &0.73 &0.47 &0.29  \\
7  &1.26 &0.88 &0.56 &0.34  \\
\vdots &\vdots &\vdots &\vdots &\vdots  \\ \hline\rule{0pt}{2.3ex}
\gamma_\beta  &110.0&97.0 &75.5 &53.2 \\[2pt]
\hline
\end{array}
\end{math}}
\label{asymrange}
\caption{First six perturbative coefficients in the expansions of
  the $\beta$-function in powers of the bare coupling constant $g_B$,
 divided by their asymptotic large-order estimates $(- \alpha )^kk!k^{ \beta _
 \beta }$. The ratios increase quite slowly towards
  the theoretically predicted normalization constant $\gamma_\beta $
in the asymptotic regime.}
\end{table}
\begin{equation}
  \label{eq:defbeta}
  {\beta}_k/\beta ^{\rm as}_k\equiv \beta_k/k!(- \alpha )^kk^{\bOj }.
\end{equation}
The first six approach their large-order limits
quite slowly.
For this reason we prefer to
adapt $ \sigma $ not from  $ \alpha $
by Eq.~(\ref{@thischoice}), but by an optimization
of the convergence. Since the reexpanded series converges for fixed values of
$\delta$ and $\sigma$ it is reasonable to determine these parameters by
searching for a point of least dependence in largest available order $L$.
This is done by imposing the conditions
\begin{equation}
  \frac{d \kappa _L}{d\sigma}=0\;\; \mathrm{and} \;\;
  \frac{d^2 \kappa _L}{d\sigma^2}=0
\end{equation}
to determine {\em both} parameters $\delta,\;\sigma$, where $\kappa_L$
denotes the $L$th approximation to
any exponent $\gamma,\nu$ or $\eta$.
In accordance with the discussions in section 2.1.2 this procedure provides
a value of $\sigma$ which is smaller than that given by (\ref{@thischoice}).

After trying out a few choices,
we have given the parameters $\beta$ and
$\rho$ the fixed values $1$ and $10$, respectively, to accelerate the
convergence.

\begin{figure}[h]
\unitlength=1mm
\begin{picture}(48.5,100)
\put(-2,52){\IncludeEpsImg{48.5mm}{48.5mm}{0.7000}{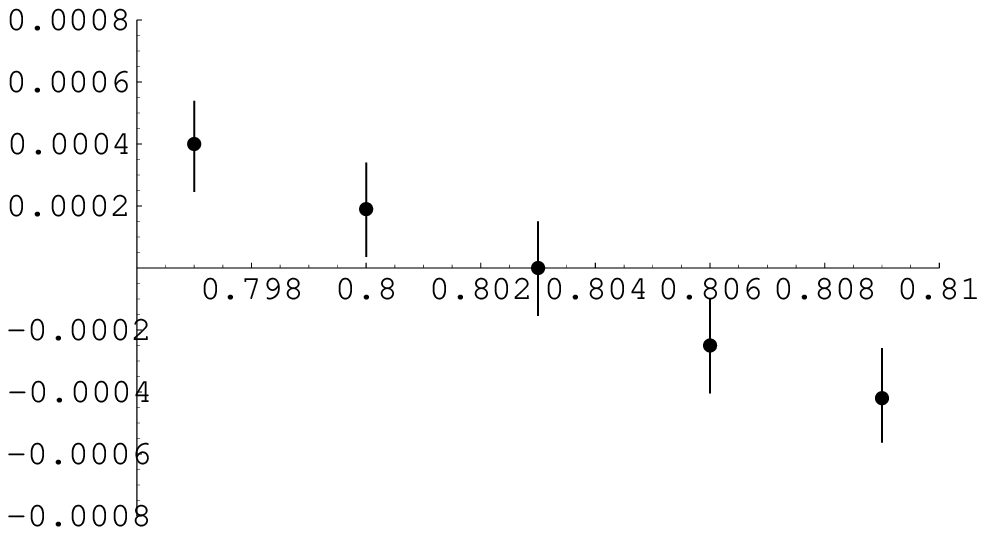}}
\put(88,52){\IncludeEpsImg{48.5mm}{48.5mm}{0.7000}{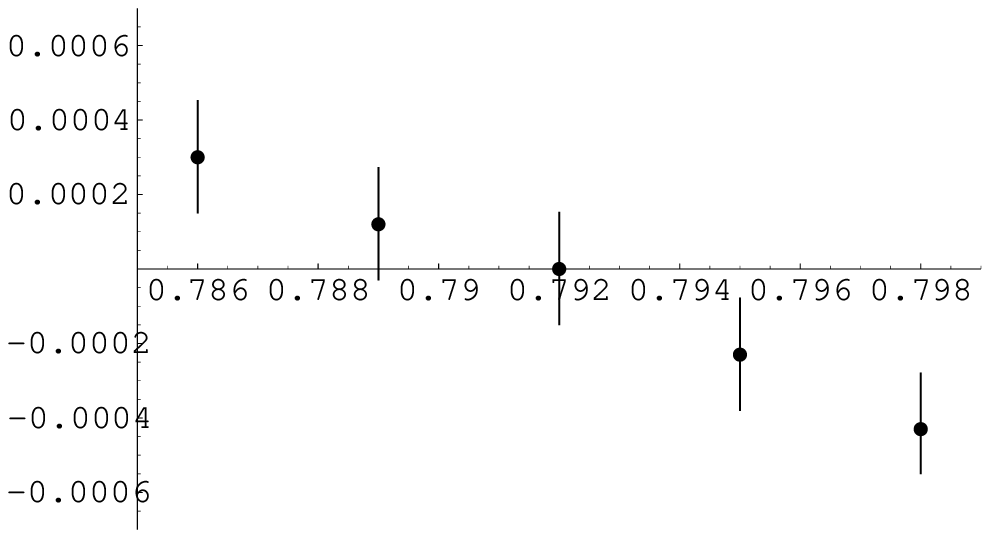}}
\put(-2,0){\IncludeEpsImg{48.5mm}{48.5mm}{0.7000}{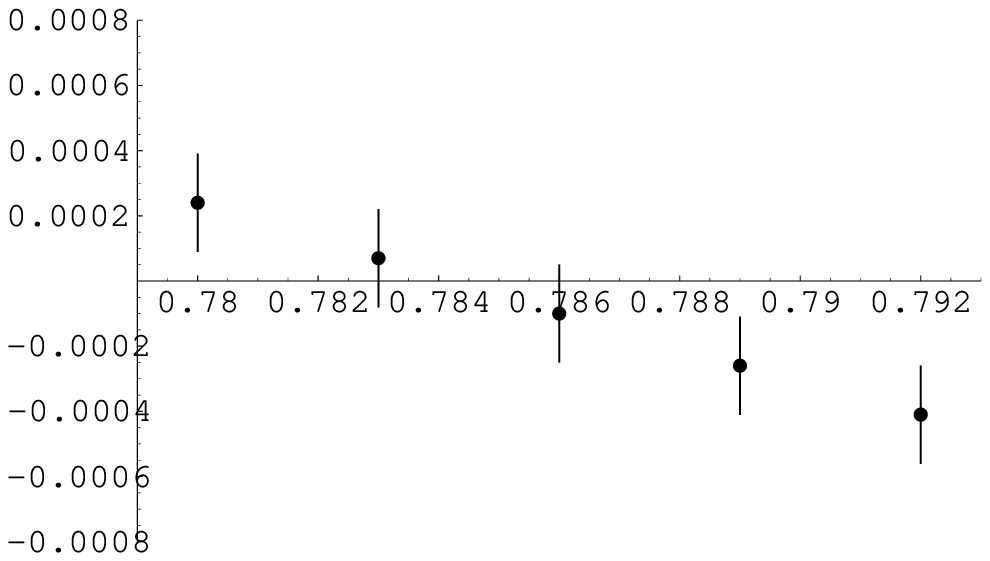}}
\put(88,0){\IncludeEpsImg{48.5mm}{48.5mm}{0.7000}{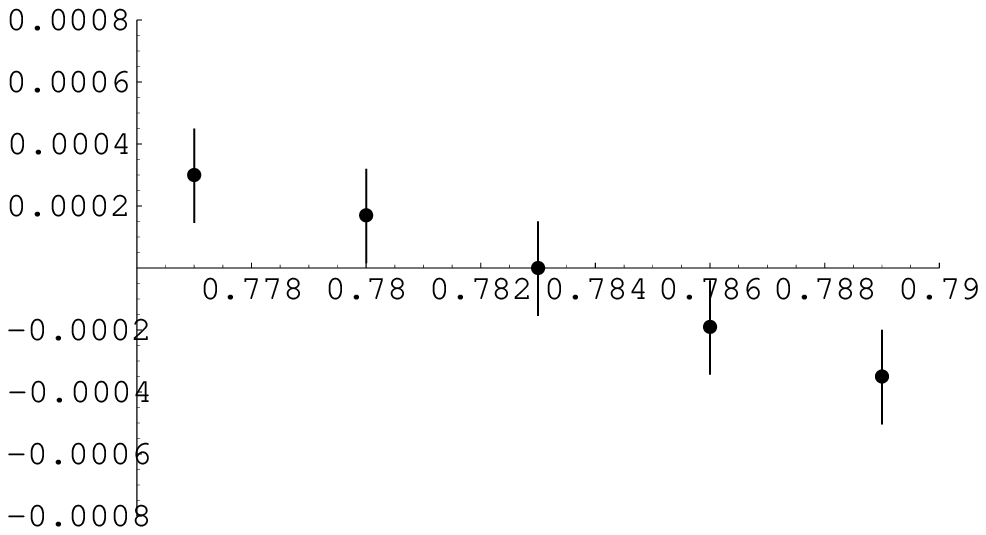}}

\put(68,66){\fsz$\omega$}
\put(11,89){\fsz$\beta^*$ }

\put(156,66){\fsz$\omega$}
\put(101,89){\fsz$\beta^*$}

\put(68,15){\fsz$\omega$}
\put(11,36){\fsz$\beta^*$}

\put(156,15){\fsz$\omega$}
\put(101,36){\fsz$\beta^*$}

\put(45,80){\fsz$n=0$}
\put(135,80){\fsz$n=1$}

\put(45,28){\fsz$n=2$}
\put(135,28){\fsz$n=3$}
\end{picture}
\caption{Plot of resummed values of $\beta^*$
  against $\omega$. The true value of $\omega$ is deduced from the condition
   $\beta^*=0$ and the errors are determined from the range of $\omega$ where
   the error bars from the resummation of $\beta^*$ intersect with the $x$-axis.}
\label{erom}
\end{figure}

The results for the critical exponents of all O($N$)-symmetries are
shown in  Figs.~\ref{fig1}--\ref{fign} and Table~3.
\begin{table}[hb]
\makebox[\textwidth]{\begin{math}
\begin{array}{|c|l|l|l|l|} \hline
\rule{0pt}{2.3ex}
n & ~~~~\gamma & ~~~~\eta &~~~~ \nu & ~~~\omega \\[2pt] \hline
\rule{0pt}{2.3ex}
0 & 1.1604[8]~\,\,(4)\;\{0.075\} &
0.0285[6]~\,(4)\;\{0.037\} & 0.5881[8]~\,(4)\;\{0.075\} & 0.803[3]~\{1\} \\
1 & 1.2403[8]~\,\,(4)\;\{0.110\} & 0.0335[6]~\,(3)\;\{0.043\} & 0.6303[8]~\,(4)\;\{0.065\}
& 0.792[3]~\{1\} \\
2 & 1.3164[8]~\,\,(5)\;\{0.033\} & 0.0349[8]~\,(5)\;\{0.042\} & 0.6704[7]~\,(4)\;\{0.098\}
& 0.784[3]~\{1\} \\
3 & 1.3882[10]\,(7)\;\{0.210\} & 0.0350[8]~\,(5)\;\{0.043\} & 0.7062[7]~\,(4)\;\{0.110\} &
0.783[3]~\{1\} \\[2pt] \hline
\end{array}
\end{math}}
\label{exp}
\caption{Critical exponents of the O($N$)-symmetric $\phi^4$-theory from our
  new resummation method. The numbers in square brackets indicate the total
  errors. They arise form the error of the resummation at fixed values of
  $\omega$ indicated in parentheses,
  and the errors coming from the inaccurate
knowledge of $ \omega $.
The former
  are estimated from the scattering of the
  approximants around the graphically determined large-$L$ limit, the
latter follow from
the errors in $ \omega $ and the
derivatives of the critical exponents
with respect to changes of $ \omega $ indicated
in the curly brackets.}
\end{table}%

The total error is indicated in the square brackets. It is deduced from
the error of resummation of the critical exponent at a
fixed value of $\omega$ indicated in the parentheses,
and from the error
$\Delta \omega$ of $\omega$, using the derivative of the exponent
with respect to $\omega$ given in curly brackets.   Symbolically, the
 relation between these errors is
\begin{equation}
  [\ldots]=(\ldots)+\Delta\omega\{\ldots\}.
\end{equation}
The accuracy of our results can be judged by comparison with
the
most accurately measured critical exponent
 $ \alpha $ parametrizing the divergence
of the specific heat of superfluid helium
at the $ \lambda $-transition by $|T_c-T|^{- \alpha }$.
By going into a vicinity of the critical temperature
with  $ \Delta T\approx10^{-8}$ K,
a recent satellite experiment
has provided us with
the value \cite{rLipa}
\begin{equation}
\alpha =-0.01285\pm0.00038.
\label{@expnum}\end{equation}
Our value for $ \nu $ in Table~3 is
\begin{equation}
 \nu  =-0.6704\pm0.007
\label{}\end{equation}
and yields via the scaling relation
$ \alpha =2-3 \nu $:
\begin{equation}
\alpha =-0.0112\pm0.0021,
\label{@ourresja}\end{equation}
in good agreement with the experimental number (\ref{@expnum}).
A comparison with other experiments and theories
is shown in Fig.~\ref{@nu2my}, showing that
our result is among the more accurate ones.

A remark is necessary concerning the errors quoted
in this paper.
We do not know how to estimating
precisely
the errors which can appear in
an involved
numerical approximation scheme such as the one presented here.
Our estimates are based on the range of critical exponents
which can be reached by reasonably modifying the parameters in
the calculations. What may be considered as reasonable is a
somewhat subjective procedure.
As such, our  error estimates follow the
rule of maximal optimism, and are probably underestimated.
This is, however, not uncommon in
resummations of divergent power series
of critical exponents.

~\\~\\

\begin{figure}[h]
\unitlength=1mm
\begin{picture}(48.5,100)
\put(-2,52){\IncludeEpsImg{48.5mm}{48.5mm}{0.7000}{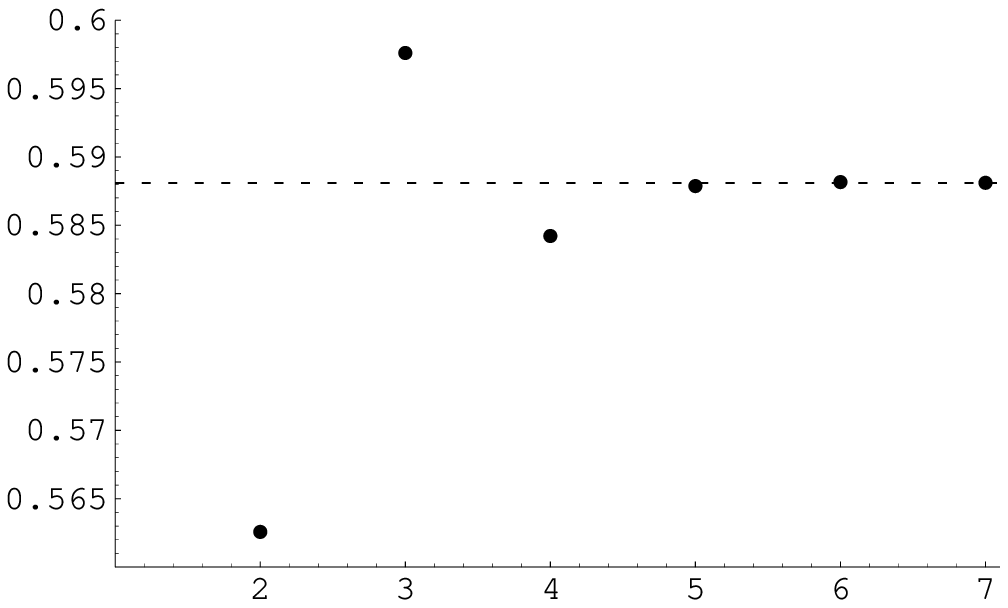}}
\put(88,52){\IncludeEpsImg{48.5mm}{48.5mm}{0.7000}{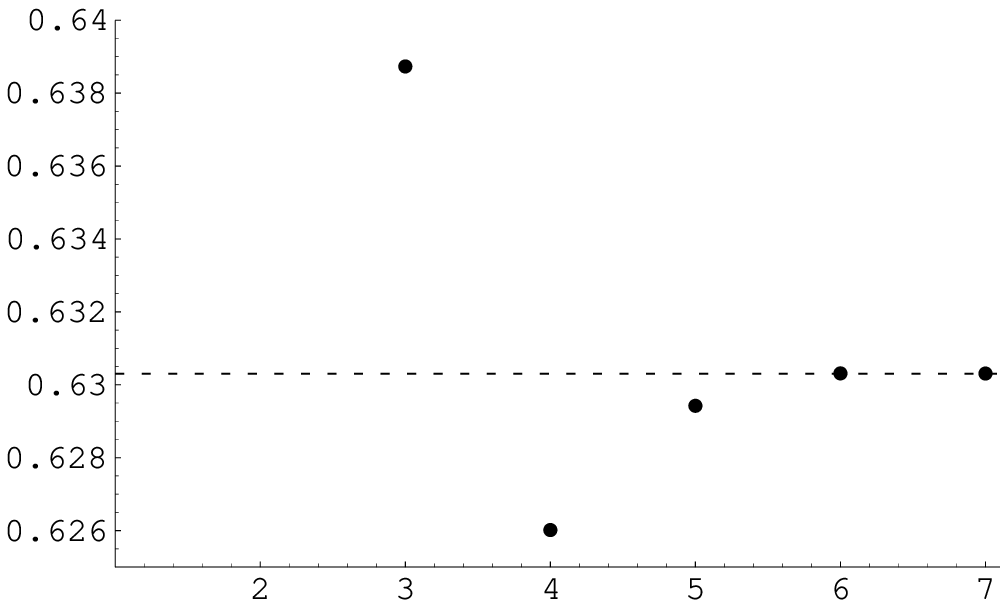}}
\put(-1,0){\IncludeEpsImg{48.5mm}{48.5mm}{0.7000}{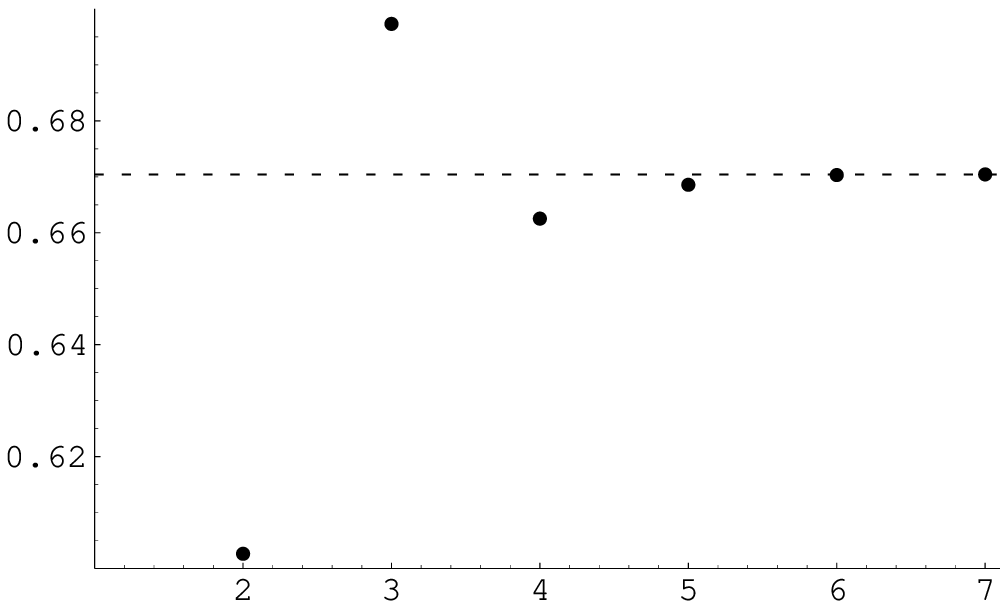}}
\put(90,0){\IncludeEpsImg{48.5mm}{48.5mm}{0.7000}{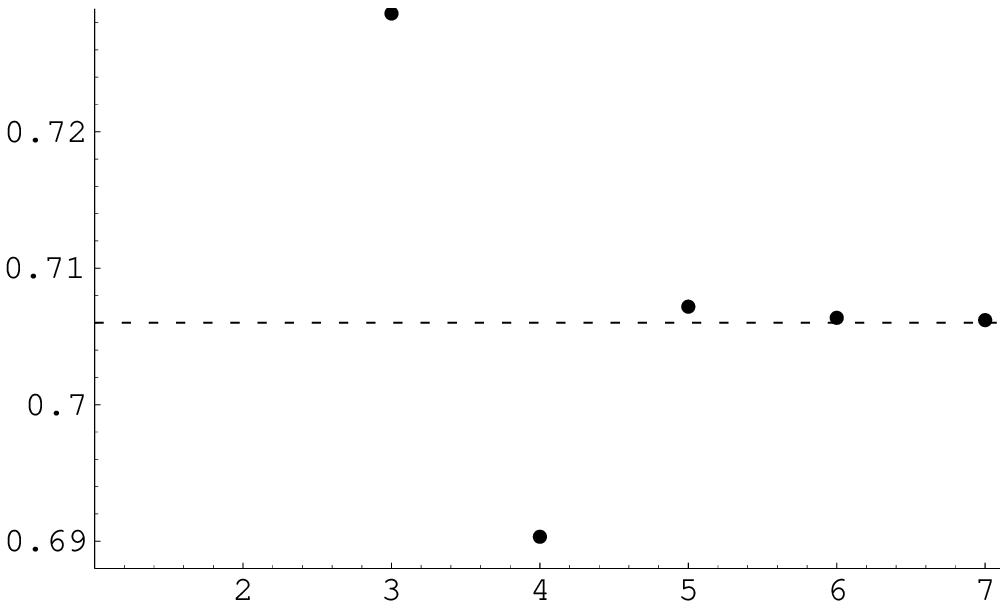}}

\put(10,84){\fsz$0.5881$}
\put(11,94){\fsz$\nu_L$}

\put(100,70){\fsz$0.6303$}
\put(101,94){\fsz$\nu_L$}

\put(10,32){\fsz$0.6704$}
\put(11,41){\fsz$\nu_L$}

\put(100,22){\fsz$0.7062$}

\put(100,37){\fsz$\nu_L$}

\put(35,68){\fsz$n=0$}
\put(135,78){\fsz$n=1$}
\put(35,16){\fsz$n=2$}
\put(135,26){\fsz$n=3$}
\end{picture}
\caption{Convergence of the approximations $\nu_L$
to the critical exponent $ \nu $ for different values of  $N$.}
\end{figure}

\begin{figure}[h]

\unitlength=1mm
\begin{picture}(48.5,100)
\put(-2,52){\IncludeEpsImg{48.5mm}{48.5mm}{0.7000}{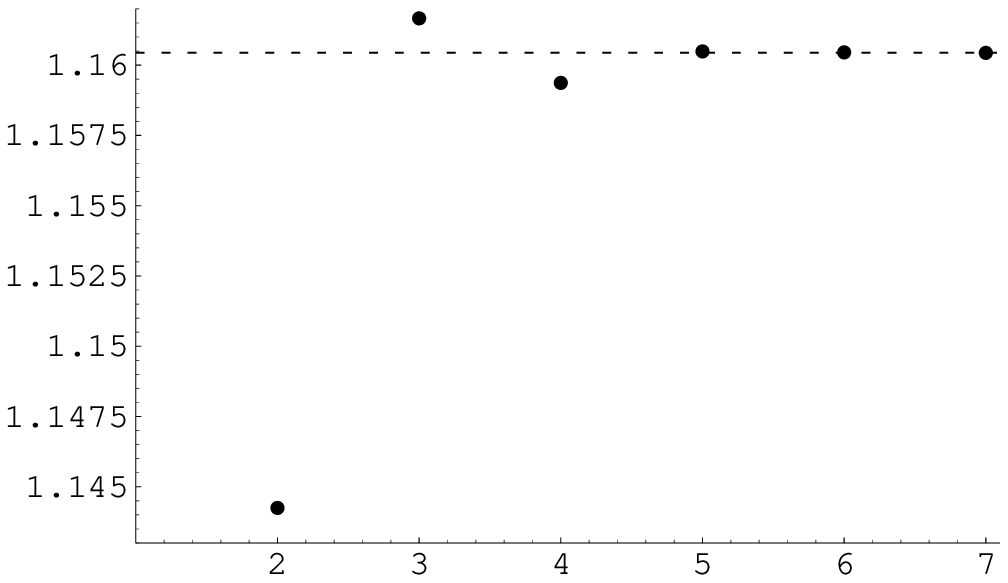}}
\put(88,52){\IncludeEpsImg{48.5mm}{48.5mm}{0.7000}{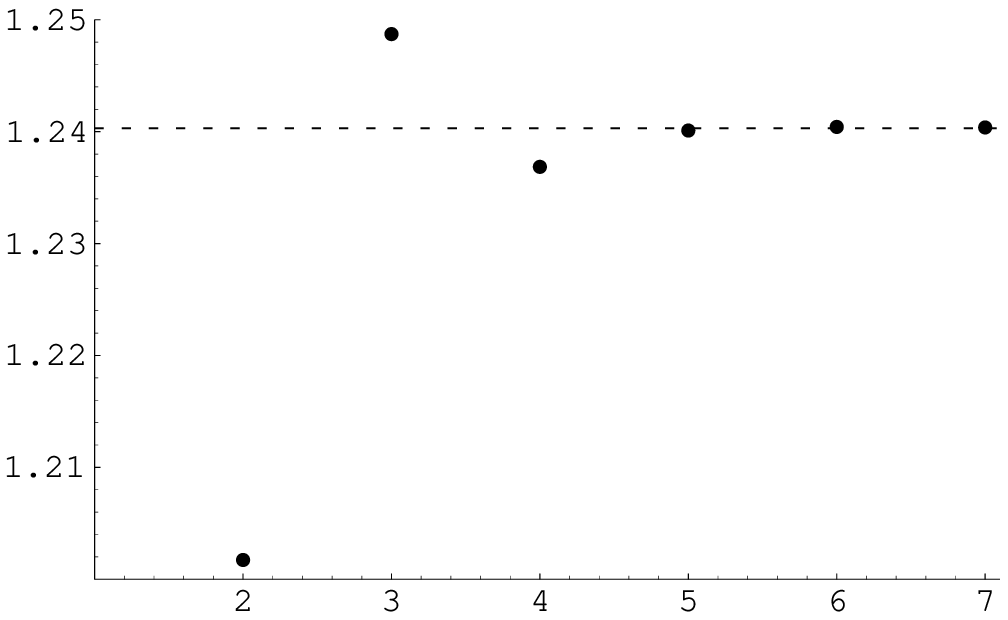}}
\put(-.5,0){\IncludeEpsImg{48.5mm}{48.5mm}{0.7000}{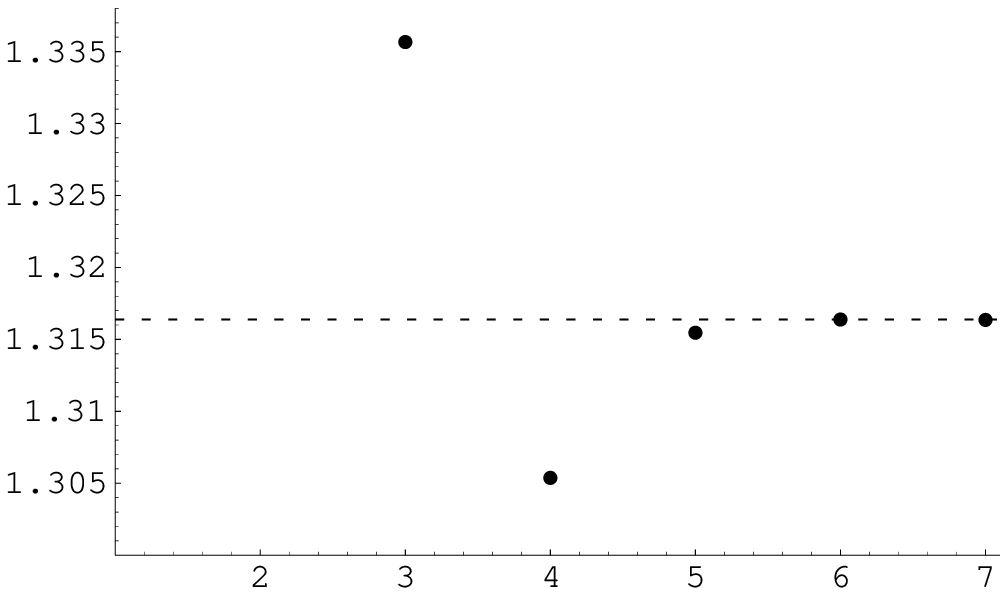}}
\put(88,0){\IncludeEpsImg{48.5mm}{48.5mm}{0.7000}{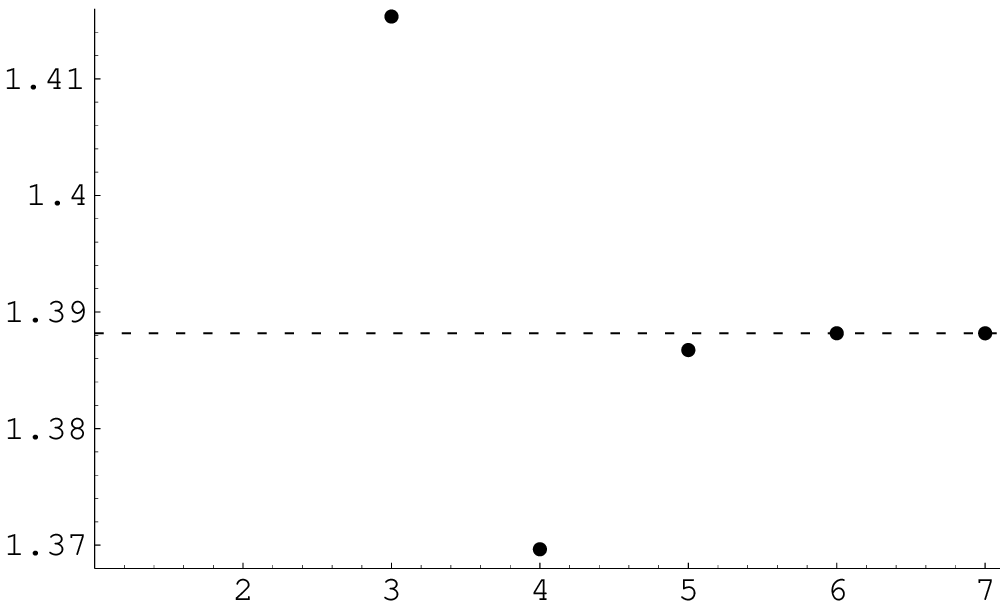}}

\put(45,92){\fsz$1.1604$}
\put(11,94){\fsz$\gamma_L$}

\put(135,90){\fsz$1.2403$}
\put(101,94){\fsz$\gamma_L$}

\put(45,22){\fsz$1.3164$}
\put(11,41){\fsz$\gamma_L$}

\put(135,22){\fsz$1.3882$}
\put(100,37){\fsz$\gamma_L$}

\put(35,68){\fsz$n=0$}
\put(125,68){\fsz$n=1$}

\put(20,16){\fsz$n=2$}
\put(110,16){\fsz$n=3$}

\end{picture}
\caption{Convergence of the approximations $ \gamma _L$
to the critical exponent $  \gamma  $ for different values of  $N$.}
\end{figure}

\begin{figure}[h]
\unitlength=1mm
\begin{picture}(48.5,100)
\put(-2,52){\IncludeEpsImg{48.5mm}{48.5mm}{0.7000}{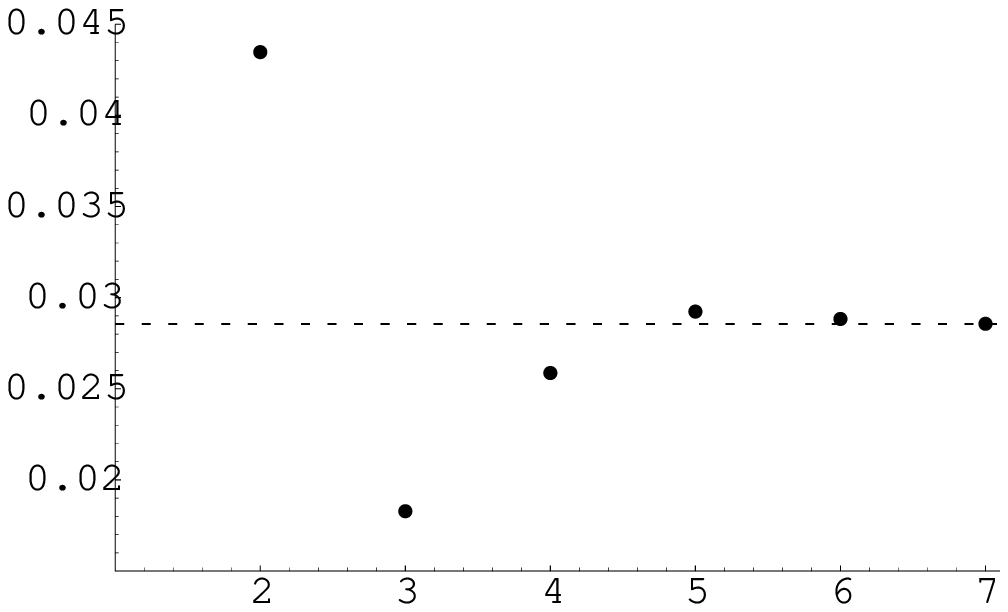}}
\put(88,52){\IncludeEpsImg{48.5mm}{48.5mm}{0.7000}{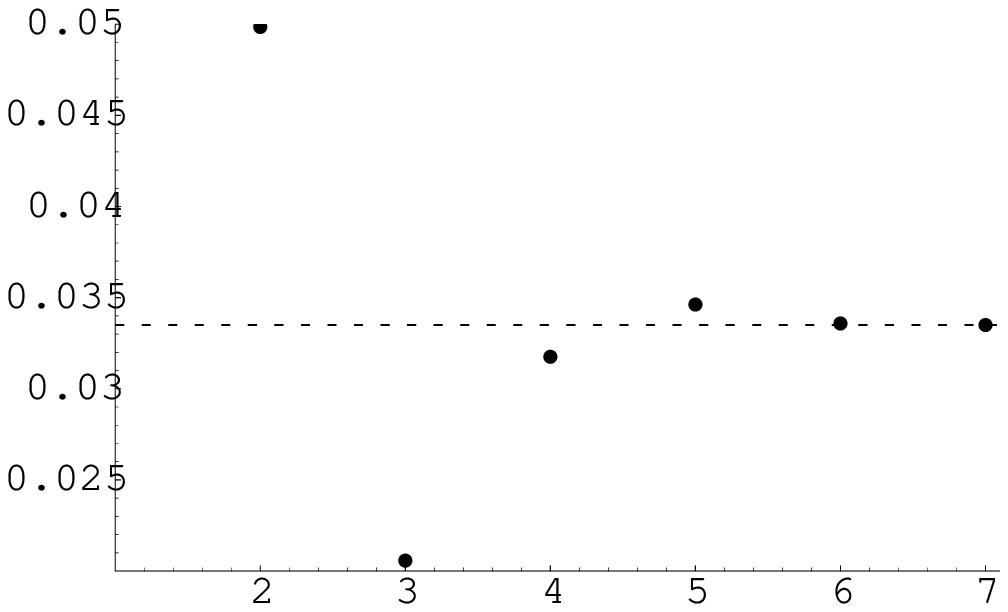}}
\put(-2,0){\IncludeEpsImg{48.5mm}{48.5mm}{0.7000}{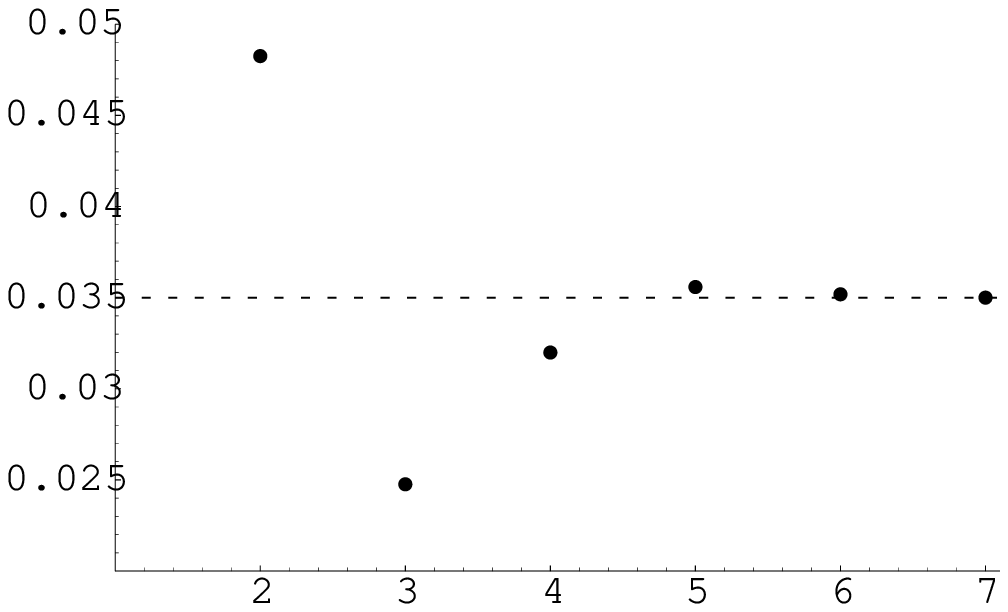}}
\put(88,0){\IncludeEpsImg{48.5mm}{48.5mm}{0.7000}{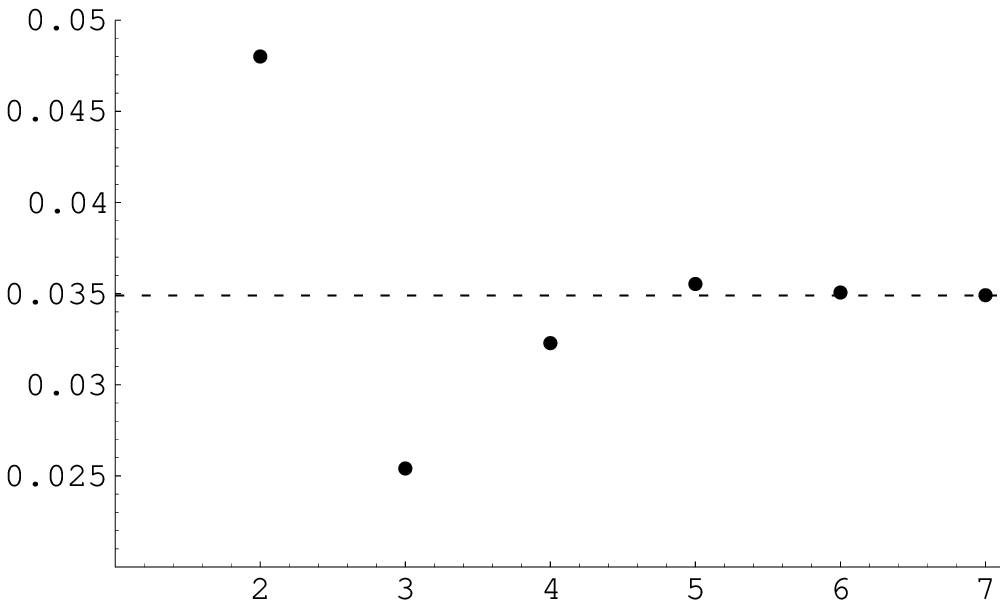}}

\put(15,74){\fsz$0.0285$}
\put(9,89){\fsz$\eta_L$}

\put(105,74){\fsz$0.0335$}
\put(99,89){\fsz$\eta_L$}

\put(15,24){\fsz$0.0350$}
\put(9,36){\fsz$\eta_L$}

\put(105,24){\fsz$0.0349$}
\put(99,36){\fsz$\eta_L$}

\put(45,68){\fsz$n=0$}
\put(135,68){\fsz$n=1$}

\put(45,16){\fsz$n=2$}
\put(135,16){\fsz$n=3$}
\end{picture}
\caption{Convergence of the approximations $ \eta_L$
to the critical exponent $  \eta  $ for different values of  $N$.}
\label{fign}
\end{figure}

\begin{figure}[tbhp]
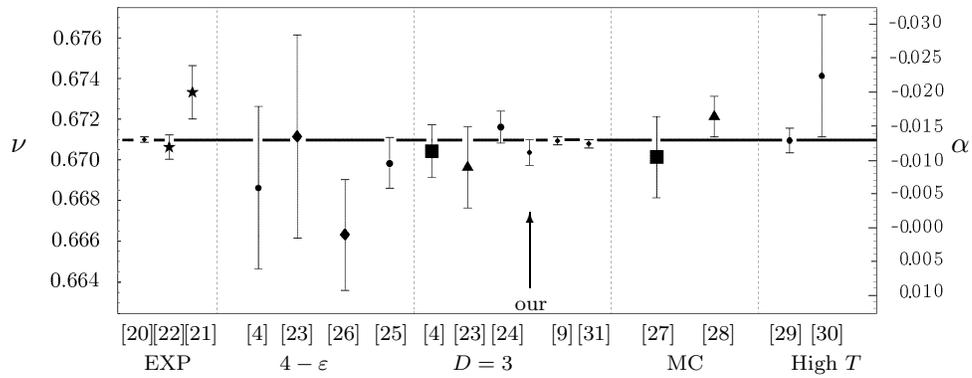

\vspace{2cm}
\input  nu2my.tps
\caption[]{Comparison of our result (\ref{@ourresja})
for critical exponents $ \alpha $
of superfluid helium with experiments and other theories.}
\label{@nu2my}\end{figure}


\begin{thebibliography}{99}
\bibitem{nickel} D. B. Murray and B. G. Nickel, unpublished
\bibitem{KNS}H. Kleinert, J. Neu, V. Schulte-Frohlinde,
K.G. Chetyrkin, and  S.A. Larin, Phys. Lett.{\bf B 272}, 39 (1991)
\bibitem{KSF}
H. Kleinert and V. Schulte-Frohlinde, Phys. Lett. {\bf B 342}, 284 (1995)
\bibitem{GZ}
R.~Guida and J.~Zinn-Justin,
{\em Critical exponents of the $N$-vector model\/},
J. Phys. A {\bf 31}, 8130 (1998)
(cond-mat/9803240)
\bibitem{PI}
For a detailed description of this theory
in quantum mechanics see the textbook\\
H. Kleinert, {\em Path Integrals in Quantum Mechanics
    Statistics and Polymer Physics}, World Scientific, Singapore 1995
 (http://www.physik.fu-berlin.de/\~{}kleinert/b3).
The extension to quantum field theory is described in Chapter 19 of the textbook
H. Kleinert and V. Schulte-Frohlinde, {\em Critical Properties of $\phi^4$-Theories},
World Scientific, Singapore 2000.
 (http://www.physik.fu-berlin.de/\~{}kleinert/b8).
\\
There were two main predecessors
to variational perturbation theory coming from two different directions.
From the mathematical side, the seminal paper was
\\
{R.~Seznec and J.~Zinn-Justin}, J.~Math.~Phys.~{\bf 20}, 1398 (1979).\\
From the physical side, inspiration came from \\
 R.P Feynman and H. Kleinert,
  Phys. Rev. A {\bf 34}, 5080 (1986)
 (http://www.physik.fu-berlin.de/\~{}kleinert/159),\\
and its
 systematic extension in\\
{H. Kleinert},
Phys. Lett. A {\bf 173}, 332 (1993)
(quant-ph/9511020).
 For the contributions of numerous other authors see
Notes and References of Chapters 5 and~17 in the first textbook.
A small excerpt is given below.

The crucial breakthrough which opened up the
previous quantum mechanical variational approaches
to quantum field theory came in three steps.
First, still in quantum mechancis, by exploiting
previously
unused even
approximants which do not have an extremum, as explained
in Chapter 5 of
the textbook.
For applications to quantum field theory, two more ingredients
were important, as pointed out in
 Refs.~\cite{PI3} and \cite{seven}:
the determination of the exponent $ \omega $
by the leading power behavior in the strong-coupling limit, and
 an
extrapolation procedure to infinite order on the basis of the
theoretically known analytic order dependence.
These developments were essential
in obtaining accurate critical exponents
rivaling the
powerful combination of renormalization group
and Borel-type resummation methods.
Variational perturbation theory also  yields
directly $ \epsilon $-expansions for the critical exponents
without
the renormalization group formalism, as shown in
 Ref.~\cite{PIep}.\\
A selected list of important contributions is\\
{T.~Barnes and G.I.~Ghandour}, Phys.~Rev.~D {\bf  22},
  924 (1980);\\
{B.S.~Shaverdyan} and
{A.G.~Usherveridze},
 Phys.~Lett.~B {\bf 123}, 316 (1983);\\
{P.M.~Stevenson}, Phys.~Rev.~D {\bf  30}, 1712 (1985);\\
D {\bf  32}, 1389 (1985);\\
{P.M.~Stevenson} and
{ R.~Tarrach}, Phys.~Lett.~B {\bf 176}, 436 (1986);\\
{A.~Okopinska}, Phys.~Rev. D {\bf  35}, 1835 (1987);\\
D {\bf  36}, 2415 (1987);\\
{W.~Namgung},
{P.M.~Stevenson}, and
{J.F.~Reed},
 Z.~Phys.~C {\bf  45}, 47 (1989);\\
{U.~Ritschel}, Phys.~Lett.~B {\bf 227}, 44 (1989);\\
    Z.~Phys.~C {\bf  51}, 469 (1991);\\
{M.H.~Thoma}, Z.~Phys.~C {\bf  44}, 343 (1991);\\
{I.~Stancu} and
{P.M.~Stevenson},
  Phys.~Rev.~D {\bf  42}, 2710 (1991);\\
{R.~Tarrach}, Phys.~Lett.~B {\bf 262}, 294 (1991);\\
{H.~Haugerud} and
{F.~Raunda}, Phys.~Rev.~D {\bf  43},
  2736 (1991);\\
{A.N.~Sissakian},
{I.L.~Solivtosv}, and
  {O.Y.~Sheychenko},
  Phys.~Lett.~B {\bf 313}, 367 (1993);\\
{A.~Duncan} and
{H.F.~Jones}, Phys.~Rev.~ D {\bf  47}, 2560
    (1993);\\
For the anharmonic oscillator,
the highest accuracy in the strong-coupling
limit was reached
 with exponentially fast convergence in Ref.~\cite{JK}.
  That paper contains references to earlier less
accurate
calculations of strong-coupling expansion coefficients
from weak-coupling perturbation theory, in particular\\
F.M. Fern\'{a}ndez and R. Guardiola, J. Phys. A {\bf26}, 7169 (1993);\\
F.M. Fern\'{a}ndez, Phys. Lett. A {\bf166}, 173 (1992);\\
R. Guardiola, M.A. Sol\'{\i}s, and J. Ros, Nuovo Cimento B {\bf 107},
713 (1992).
A.V. Turbiner and A.G. Ushveridze, J. Math. Phys. {\bf 29}, 2053 (1988);\\
B. Bonnier, M. Hontebeyrie, and E.H. Ticembal, J. Math. Phys. {\bf 26}, 3048
(1985);\\ Those works were unable to extract the exponential law of convergence
 from their data.
This was shown
to be related to the convergence radius of the
strong-coupling expansion by\\
 H.~Kleinert and W.~Janke,
Phys. Lett. A {\bf 206}, 283 (1995) (quant-ph/9509005);\\
and in the last paper in Ref.~\cite{z1112}.\\
Predecessors of these works
which did
not yet explain
the
exponentially fast convergence in the strong-couplings
limit
are\\
I.R.C. Buckley, A. Duncan, and H.F. Jones, Phys. Rev. D {\bf  47}, 2554 (1993);\\
C.M. Bender, A. Duncan, and H.F. Jones, Phys. Rev. D {\bf  49}, 4219 (1994);\\
A. Duncan and H.F. Jones, Phys. Rev. D {\bf  47}, 2560 (1993);\\
C. Arvanitis, H.F. Jones, and C.S. Parker,
Phys. Rev. D {\bf  52}, 3704 (1995) (hep-th/9502386);\\
R. Guida, K. Konishi, and H. Suzuki,
Annals Phys. {\bf 241}, 152 (1995) (hep-th/9407027).







\bibitem{scth}
 H.~Kleinert,
      Phys.~Lett.~A {\bf 207}, 133 (1995) (quant-ph/9507005).
\bibitem{z1112}
R.~Guida, K.~Konishi and H.~Suzuki, Ann.~Phys.~{\bf 241}, 152 (1995)
(hep-th/9407027);
 Annals Phys.~{\bf 249}, 109
(1996) (hep-th/9505084).
\bibitem{PI3}H. Kleinert, Phys. Rev. D {\bf 57}, 2264 (1998)
(www.physik.fu-berlin.de/\~{}kleinert/257);
Addendum:
ibid. D {\bf 58}, 1077 (1998) (cond-mat/9803268).
\bibitem{seven}
H. Kleinert,
Phys. Rev. D {\bf 60}, 085001 (1999) (hep-th/9812197);
{\em Theory and Satellite Experiment
on Critical Exponent alpha of Specific Heat in
Superfluid Helium\/}
(cond-mat/9906107).
\bibitem{PIep}H. Kleinert, Phys. Lett.  B {\bf 434}, 74 (1998) (cond-mat/9801167);
ibid. B {\bf 463}, 69 (1999) (cond-mat/990635).
\bibitem{parisi2} G. Parisi, E. Brezin, J. Stat. Phys. {\bf 19}, 269
  (1978), G. Parisi J. Stat. Phys. {\bfseries 23}, 49 (1980)
\bibitem{ch17}
For an introduction see, for instance, Chapter 17 in the textbook \cite{PI}.
\bibitem{parisi} G. Parisi, Phys. Lett. B {\bfseries  69 }, 329 (1977)
\bibitem{z13}
 J.J.~Loeffel, in {\em Large-Order Behaviour of Perturbation Theory\/},
ed. by
 J.C.~Le Guillou and J. Zinn-Justin, North-Holland, Amsterdam, 1990.
\\
 J.C.~Le Guillou and J. Zinn-Justin, J.~Phys.~Lett.~(Paris) {\bf 46}, L137
(1985); J.~Phys.~(Paris) {\bf 48}, 19 (1987); ibid. {\bf 50}, 1365 (1987).
\bibitem{tr}This transformation has never been investigated
in the literature, although it is contained in a class of
quite general mathematical transformations
introduced in the textbook of
Hardy, {\em Divergent Series } (Oxford University Press, Oxford 1949
in the context of
{\em moment constant methods}.
These comprise
transformation $ B(y)=\sum f_k y^k/\mu_k $,
where the $\mu_k$ are given by a Stieltjes
integral $\mu_k = \int_0^\infty x^k d\chi(x) $ and $\chi$ is a bounded and increasing
function of $x$ guaranteeing the convergence of the Stieltjes integral.
This definition
includes our transformation for the somewhat compicated
choice $ d\chi(x) =({\Gamma(\betaj )}/{2\pi  i})x^{-s-1}\oint_Cdt \; e^{t+x^\omega t^{1-\omega}}t^{s(1-1/\omega )-\beta_0} \; dx. $

\bibitem{bender} C. M. Bender, T. T. Wu, Phys. Rev. {\bfseries 184} 1231 (1969)
\bibitem{simon} B. Simon, Ann. Phys. {\bfseries 58} 76 (1970)

\bibitem{JK}
  W. Janke and H. Kleinert,
 Phys. Rev. Lett. {\bf 75}, 2787 (1995) (quant-ph/9502019)
\bibitem{sokolov} S. A. Antonenko and A. I. Sokolov, Phys. Rev. E {\bfseries
  51}, 1894 (1995); S. A. Antonenko and A.I. Sokolov, Phys. Rev. B {\bfseries
  49}, 15901 (1984).



\bibitem{rLipa}J.A. Lipa, D.R. Swanson, J. Nissen,
T.C.P. Chui and U.E. Israelson, {Phys. Rev. Lett.} {\bf 76}, 944 (1996).

\bibitem{ahl}
G. Ahlers, Phys. Rev. A {\bf 3}, 696 (1971);
K.H. Mueller, G. Ahlers, F. Pobell,  Phys. Rev. B {\bf 14}, 2096 (1976);

\bibitem{rgomuha}{L.S. Goldner, N. Mulders and G. Ahlers,
J. Low Temp.Phys. 93
(1992) 131.}





\bibitem{rLGZJi}J.C. Le Guillou and J. Zinn-Justin,
Phys. Rev. Lett. {\bf 39}, 95 (1977);
Phys. Rev. B {\bf 21}, 3976 (1980); J. de Phys. Lett {\bf 46}, L137 (1985).


\bibitem{MN}
The displayed value comes from Murray and Nickel in
 Ref.~\cite{nickel},
assuming a certain  fixed point coupling
constant $g^*$.

\bibitem{KVSF}
 H. Kleinert and V. Schulte-Frohlinde,
(cond-mat/9907214).




\bibitem{rPelVic}{A. Pellisetto and E. Vicari,
Nucl. Phys. B {\em 519}, 626 (1998)
(cond-mat/9711078).}



\bibitem{rJanke}{W. Janke, Phys.Lett. A148 (1990) 306.}

\bibitem{rBFMM}H.G. Ballesteros, L.A. Fernandez, V. Martin-Mayor and A. Munoz
Sudupe, Phys. Lett. B {\bf 387}, 125 (1996).

\bibitem{WOR}M. Ferer, M.A. Moore, and M. Wortis,
Phys. Rev. B {\bf 65}, 2668 (1972).


\bibitem{rBUCOM}P. Butera and M. Comi, {Phys. Rev.} B {\bf 56}, 8212 (1997)
(hep-lat/9703018).


\bibitem{alpha}H. Kleinert,
{\em
Theory and Satellite Experiment
for Critical Exponent $ \alpha $ \\ of $ \lambda $-Transition in
Superfluid Helium\/}, (cond-mat/9906107).


\end{thebibliography}
\end{document}